\documentclass[%
aps,
 prc,%
 amsmath,amssymb,
 reprint,%
 superscriptaddress,
]{revtex4-1}

\usepackage{dcolumn}
\usepackage{graphicx}
\usepackage{lineno}
\usepackage{bm}%
\usepackage[pdftex,colorlinks=true,bookmarks=false,citecolor=blue,urlcolor=blue]{hyperref}
\usepackage{upgreek}
\usepackage{titlesec} 
\usepackage{braket}

\makeatletter
\renewcommand\frontmatter@abstractwidth{\dimexpr\textwidth-0.0in\relax}
\makeatother

\begin{document}


\title{\vspace{-15mm}\fontsize{19pt}{10pt}\selectfont\textbf{Coherent quantum dynamics of systems with coupling-induced creation pathways}} 

\author{Steven D. Rogers}
\affiliation{Department of Physics and Astronomy, University of Rochester, Rochester, NY 14627}
\affiliation{Center for Coherence and Quantum Optics, University of Rochester, Rochester, NY 14627}

\author{Austin Graf}
\altaffiliation{These authors contributed equally to this work}
\affiliation{Center for Coherence and Quantum Optics, University of Rochester, Rochester, NY 14627}
\affiliation{Institute of Optics, University of Rochester, Rochester, NY 14627}

\author{Usman A. Javid}
\altaffiliation{These authors contributed equally to this work}
\affiliation{Center for Coherence and Quantum Optics, University of Rochester, Rochester, NY 14627}
\affiliation{Institute of Optics, University of Rochester, Rochester, NY 14627}

\author{Qiang Lin}
\email[Electronic mail: ]{qiang.lin@rochester.edu}
\affiliation{Center for Coherence and Quantum Optics, University of Rochester, Rochester, NY 14627}
\affiliation{Institute of Optics, University of Rochester, Rochester, NY 14627}
\affiliation{Department of Electrical and Computer Engineering, University of Rochester, Rochester, NY 14627}

\date{\today}


\begin{abstract}
\noindent{Many technologies emerging from quantum information science heavily rely upon the generation and manipulation of entangled quantum states. Here, we propose and demonstrate a new class of quantum interference phenomena that arise when states are created in and coherently converted between the propagating modes of an optical microcavity. The modal coupling introduces several new creation pathways to a nonlinear optical process within the device, which quantum mechanically interfere to drive the system between states in the time domain. The coherent conversion entangles the generated biphoton states between propagation pathways, leading to cyclically evolving path-entanglement and the manifestation of coherent oscillations in second-order temporal correlations. Furthermore, the rich device physics is harnessed to tune properties of the quantum states. In particular, we show that the strength of interference between pathways can be coherently controlled, allowing for manipulation of the degree of entanglement, which can even be entirely quenched. The states can likewise be made to flip-flop between exhibiting initially correlated or uncorrelated behavior. Based upon these observations, a proposal for extending beyond a single device to create exotic multi-photon states is also discussed.}  
\end{abstract}

\maketitle

\noindent  A worldwide effort is underway to unlock the practical and potentially transformative utilities of quantum systems\cite{OBrien09,Kurizki15,Heshami16,Acin18,Awschalom18,OBrien18,Thompson18}. If successful, a broad range of fields stand to be revolutionized, including information processing \cite{Milburn01,Pittman02,Milburn07,OBrien09,OBrien15}, simulation \cite{Aspuru-Guzik12}, communication \cite{Gisin07}, security \cite{Shor00,Gisin02}, and metrology \cite{Dowling02,Maccone04}. And as with many nascent technological revolutions, it is not immediately clear which architectures will prove most useful in realizing these developments. It is, however, without doubt that the efficacy of such systems is linked to how proficiently they can generate and manipulate quantum states and their entanglement. To this end, it is especially important that new concepts are developed which carry out these functions while remaining broadly implementable.  
  
As a result of the research interest in this area, numerous methods have been established to generate and manipulate quantum states. A particularly promising approach involves the quantum interference of multiple excitation/creation pathways, which coherently drives a system between states in the time domain. These processes have led to a diverse set of important phenomena, including governing the dynamics of electron spins in semiconductors \cite{Belykh18}, many-body oscillations in cold atoms \cite{Kuzmich12}, superconducting flux qubits in Josephson junctions \cite{Chiorescu03}, inversionless laser oscillations in atomic media \cite{Zibrov95}, and polarization entanglement between photon pairs emitted from biexcitons \cite{Ward14}, to name a few. Here, we propose and demonstrate a new class of quantum interference phenomena that result when quantum states are created in and coherently converted between propagating electromagnetic cavity modes. We realize this concept by implementing a nonlinear optical process between the coupled counter-propagating modes of a microresonator -- establishing multiple energy-level pathways for photon pair creation. In doing so, we are able to generate tunable photonic quantum states which are imparted with intrinsic time-evolving path-entanglement and exhibit coherent oscillations in their second-order temporal correlations. In particular, we show that the ability to conveniently tune properties of the optical microresonator translates to a powerful platform for manipulating the state and its entanglement properties via quantum interference. Furthermore, we demonstrate that varying the cavity photon lifetime transforms the system from producing strongly entangled quantum states of light with extremely high-contrast two-photon interference visibility, to a regime where the entanglement and oscillations are quenched and the photon statistics return to the behavior of an uncoupled system. The device may also be configured to flexibly set the probability amplitudes associated with generating photon pairs in one or the other propagation modes, thus providing a means to explore how internal symmetry affects the quantum state and entanglement.

\begin{figure}
\begin{center}
\includegraphics[scale = 0.97]{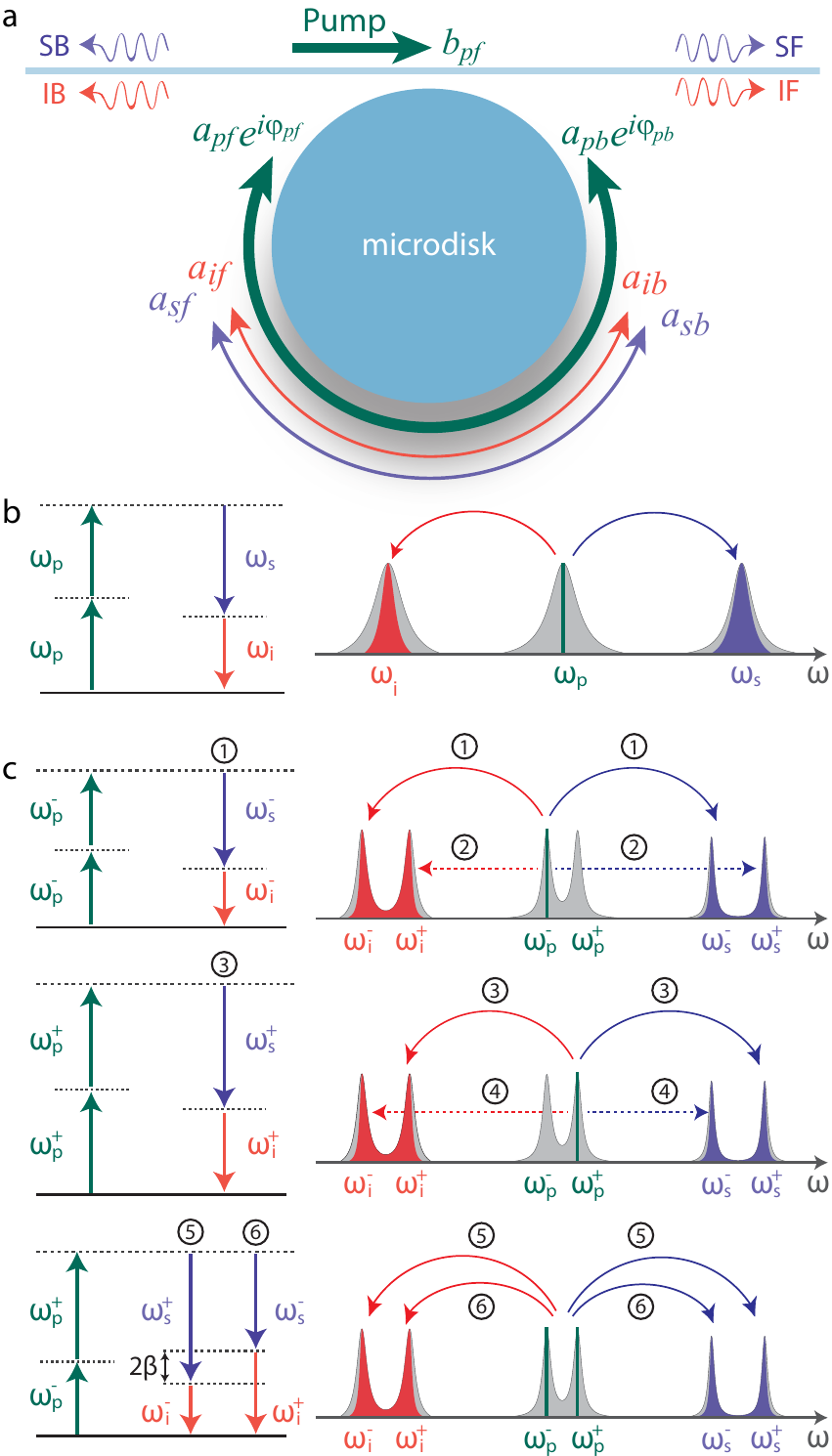}
\caption{ \small Photon pair creation between coherently coupled counter-propagating whispering-gallery modes (WGMs). (a) A forward-propagating pump wave (green) evanescently couples into the microdisk, on resonance with the pump cavity mode, building up a strong intracavity field in the forward direction, $a_{pf}$. In the presence of degenerate mode coupling, the backward-traveling intracavity pump field, $a_{pb}$, also experiences a coherent buildup. Through spontaneous four-wave mixing (SFWM), signal (blue) and idler (red) photons are created between their respective coherently coupled forward and backward intracavity modes, establishing tunable photonic quantum states with time-evolving path-entanglement. The photon pairs are coupled from the cavity into four transmitted fields: signal forward (SF), signal backward (SB), idler forward (IF), and idler backward (IB). Hence, correlations are established between photon pairs in the four path configurations: SF-IF, SF-IB, SB-IF, SB-IB. (b) Energy diagram and spectral emission profiles for SFWM in an uncoupled optical microresonator. (c) Energy diagrams and spectral emission profiles for SFWM between coupled counter-propagating WGMs. Green lines indicate which pairs of pump photons are annihilated for the various creation pathways. The dashed creation pathways are only relevant for extremely short timescales, as permitted by the uncertainty principle. }
\label{Fig1}
\end{center}
\end{figure}

We implement the concept within a whispering-gallery mode (WGM) microresonator that supports spontaneous four-wave mixing (SFWM), a $\chi^{(3)}$ nonlinear optical process \cite{Boyd08}, between three interacting cavity modes (see Fig.~\ref{Fig1}). As shown in Fig.~\ref{Fig1}(a),(b), photons are coupled from a forward-propagating pump into the pump (p) mode alone, but may spontaneously scatter into the adjacent signal (s) and idler (i) modes, in accordance with energy conservation. When weakly pumped, this vacuum-seeded nonlinear wave mixing process produces correlated bipartite states \cite{Sharping06,Leuchs16}. 

Rotationally symmetric microresonators support two degenerate modes for each resonance frequency, forward (clockwise) and backward (counterclockwise) traveling, which do not exchange energy in the absence of coupling \cite{Kippenberg02}. Thus, when photons are generated inside an uncoupled microcavity they are restricted to remain in a single propagation mode and are the result of a single creation pathway (see Fig.~\ref{Fig1}(b)), precluding quantum interference. If, however, a coupling is introduced between the counter-propagating modes, then the proposed phenomena may be realized through the establishment of a coherent conversion process. In the specific implementation considered here, photons are converted between the forward (f) and backward (b) propagation modes at a rate of $\beta_m$ (m = p,s,i), which is revealed through, $H_0 = \sum_{m} \lbrace\hbar\omega_{0m}(a_{mf}^\dagger a_{mf}^{} + a_{mb}^\dagger a_{mb}^{}) - \hbar(\beta_m a_{mf}^\dagger a_{mb}^{} + \beta_m^* a_{mb}^\dagger a_{mf}^{})\rbrace$, the unperturbed Hamiltonian (see Appendix C for the complete Hamiltonian).

\begin{figure*}[ht!]
\begin{center}
\includegraphics[scale = 1]{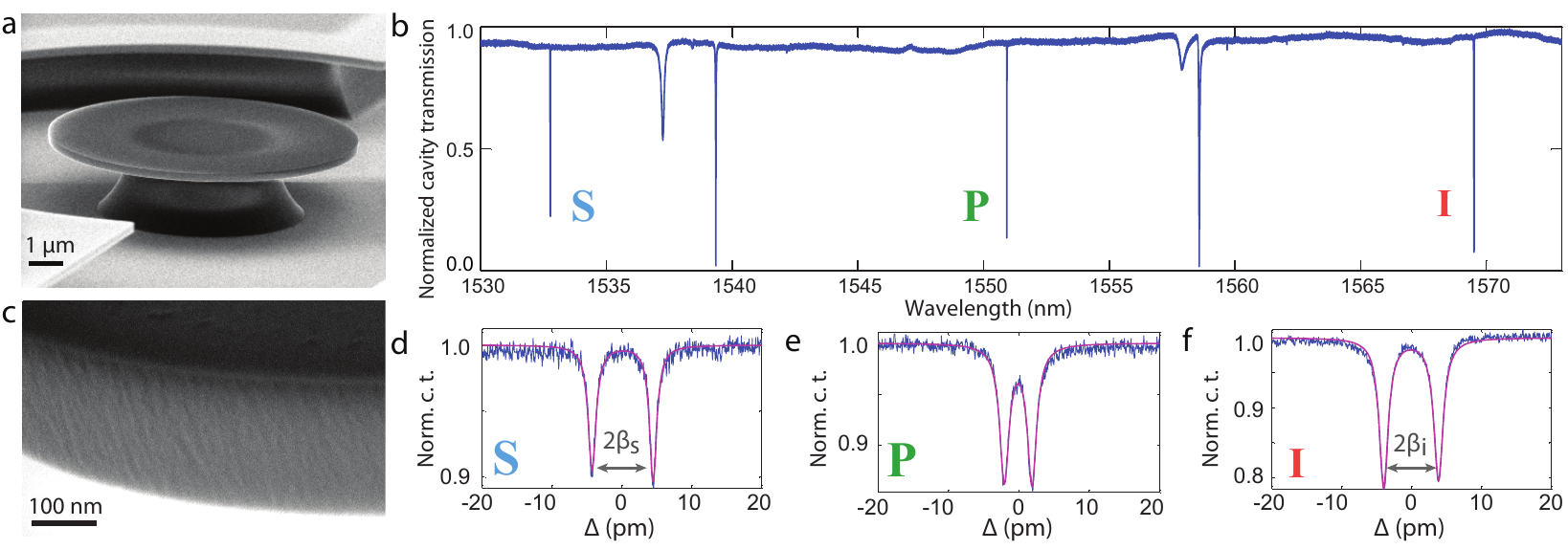}
\caption{ \small Characterization of the silicon microdisk and coupling-induced resonance splitting. (a) A scanning electron microscope (SEM) image of the silicon microdisk suspended via a silica pedestal. (b) Normalized cavity transmission spectrum with labels indicating the signal (S), pump (P), and idler (I) modes used in the cavity-enhanced spontaneous four-wave mixing (SFWM) process. (c) High magnification SEM image of the device in (a). Here, we observe the nanometer-scale surface roughness which mediates the Rayleigh-scattering-induced coupling between counter-propagating modes. (d)-(f) Detailed versions of the signal, pump, and idler doublet transmission profiles (blue), respectively, along with the fits (magenta) used to extract intrinsic optical Qs and modal coupling rates. $\Delta$: Detuning.}
\label{Fig2}
\end{center}
\end{figure*}

One of the key consequences of modal coupling is that it erases the `which-path' information from the system. For instance, when a photon exits the device it is fundamentally impossible to know whether it was originally generated in that propagation direction or converted from the counter-propagating mode. The same rule applies to its correlated partner photon, which together form the bipartite state. Thus, the probability amplitudes describing these alternative pathways add coherently, enabling quantum interference \cite{Mandel95}. To gain further insight into how modal conversion affects the system, we start by examining $H_0$. The coupling-induced energy splitting is apparent after diagonalization, which here, amounts to rotating to a standing-wave basis with eigenvectors composed of symmetric and antisymmetric combinations of traveling waves, and shifted eigenfrequencies, $\omega_{0m}^{\pm} = \omega_{0m} \pm |\beta_m|$. By applying the same transformation to the interaction Hamiltonian, we uncover how the coupling modifies the nonlinear optical processes responsible for generating the tunable photonic quantum states. In the standing-wave basis, the interaction Hamiltonian becomes $H_{int} = \frac{\hbar}{2}g(a_{s-}^\dagger a_{i-}^\dagger + a_{s+}^\dagger a_{i+}^\dagger)(a_{p-}^2 + a_{p+}^2) + \hbar g (a_{s-}^\dagger a_{i+}^\dagger + a_{s+}^\dagger a_{i-}^\dagger)a_{p-}^{} a_{p+}^{} + h.c.$, which reveals the diverse set of photon creation pathways that result from the coherent conversion between modes (see Fig.~\ref{Fig1}(c) and Appendix D).

Another critical feature of modal coupling is the effect it has on the driving field. Although a single external pump is provided, the device splits it into two counter-propagating cavity modes (see Fig.~\ref{Fig1}(a)). Now pairs of photons from either the forward or backward pump modes can participate in the nonlinear optical process. Furthermore, the modal conversion is coherent, and therefore establishes a definite phase between them. To satisfy conservation of angular momentum \cite{Boyd08}, photon pairs are only created in the co-propagating states. Thus, if we assume an undepleted classical pump and closed system, the photonic quantum states generated within the cavity will be of the form, $\ket{\psi(t=0)} = {\rm c_f}\ket{f}_s\ket{f}_i + {\rm c_b}\ket{b}_s\ket{b}_i$, where t=0 denotes the time of creation, and the complex coefficients, ${\rm c_f}$ and ${\rm c_b}$ are set by properties of the pump modes (see Appendix D). However, forward and backward are not eigenstates of the coupled cavity, and thus evolve in time, resulting in a versatile quantum state (see Appendix E),

\begingroup
\setlength{\abovedisplayskip}{3pt}
\setlength{\belowdisplayskip}{3pt}
\begin{align} \label{State}
\ket{\psi (t)} =~&\big( {\rm c_f} \cos^2 (\beta t) - {\rm c_b} \sin^2 (\beta t) \big) \ket{f}_\text{s} \ket{f}_\text{i} +  \\ &\big( {\rm c_b} \cos^2 (\beta t) - {\rm c_f} \sin^2 (\beta t) \big) \ket{b}_\text{s} \ket{b}_\text{i} + \nonumber \\ & i \left( {\rm c_f}+{\rm c_b} \right)\sin(2 \beta t) \big( \ket{f}_\text{s} \ket{b}_\text{i} + \ket{b}_\text{s} \ket{f}_\text{i}  \big), \nonumber 
\end{align} 
\endgroup

\noindent where we have assumed equal coupling rates for the pump, signal and idler modes. In the case that ${\rm c_f} = {\rm c_b}$, then we clearly see the evolution through different Bell states inside the cavity. For instance, at t=0, the state is composed of maximally entangled co-propagating photons, $\ket{\psi(t=0)} \propto \ket{f}_s\ket{f}_i + \ket{b}_s\ket{b}_i$, whereas at t = $\pi/(4\beta)$, the state is composed of maximally entangled counter-propagating photons, $\ket{\psi(t=\pi/(4\beta))} \propto \ket{f}_s\ket{b}_i + \ket{b}_s\ket{f}_i$.


\section*{Device realization}
There are numerous systems that may be used to achieve photon generation between coupled counter-propagating modes. Here, we have chosen to demonstrate this phenomenon in a high-Q silicon microdisk. In recent years, there has been interest in silicon microresonators as chip-scale sources, because they can produce ultra-pure photon pairs with high spectral brightness, strong temporal correlations, and emission wavelengths in the telecommunications band \cite{Clemmen09,Davanco12,Azzini12,Engin13,Silverstone15,Grassani15,Reimer16,Rogers16}. Additionally, they may be fabricated using complementary metal-oxide-semiconductor (CMOS) compatible processes, indicating the possibility of mass manufacturing \cite{Popovic15}.

\begin{figure}
\begin{center}
\includegraphics[width=1\columnwidth]{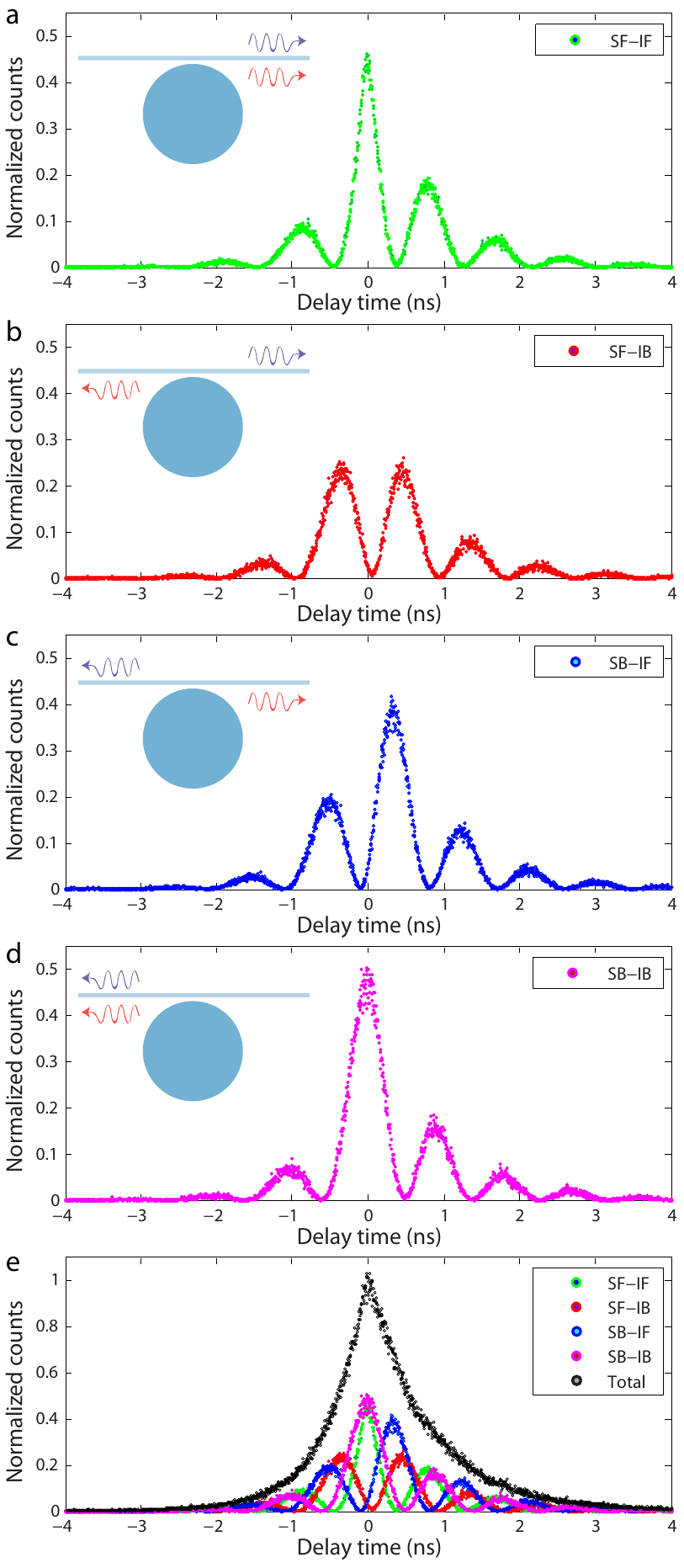}
\caption{ \small Coherent oscillations in pair correlations via quantum interference of multiple creation pathways. Cross-correlation waveforms (without background subtraction), measured between (a) signal forward - idler forward (SF-IF), (b) signal forward - idler backward (SF-IB), (c) signal backward - idler forward (SB-IF), and (d) signal backward - idler backward (SB-IB). (e) All pair correlations superposed on a single delay-time axis with the decay envelope in black resulting from the numeric sum of coincidence counts at each time bin. All data are normalized to the peak of the decay envelope. Insets depict path configuration.}
\label{Fig3}
\end{center}
\end{figure}

A scanning electron microscope (SEM) image of our suspended silicon microdisk, with a radius of approximately 4.5~$\mu$m and thickness of 260 nm, is shown in Fig.~\ref{Fig2}(a). In Fig.~\ref{Fig2}(b) we plot the normalized cavity transmission which exhibits multiple quasi-transverse-magnetic (quasi-TM) mode families, with the pump (p) at $\lambda_p = 1550.6$ nm, signal (s) at $\lambda_s = 1532.5$ nm, and idler (i) at $\lambda_i = 1569.2$ nm. Frequency matching among the interacting cavity modes is achieved by engineering the group-velocity dispersion of the device, consequently enabling efficient SFWM. The nanometer-scale roughness at the surface of the microdisk (see Fig.~\ref{Fig2}(c)) mediates Rayleigh scattering between the traveling modes, leading to a coherent coupling \cite{Mazzei07}, as is evidenced by the formation of doublets in the transmission profiles (see Fig.~\ref{Fig2}(d)-(f)). The intrinsic optical Qs for the pump, signal, and idler resonances are extracted from fits of the doublets in Fig.~\ref{Fig2}(d)-(f), and are respectively found to be, $Q_{0p} = 1.27 \times 10^6$, $Q_{0s} = 1.32 \times 10^6$, and $Q_{0i} = 1.15 \times 10^6$. Additionally, the doublet splittings for the signal and idler modes are respectively found to be $2\beta_s = 1.11$ GHz and $2\beta_i = 0.97$ GHz. The modes exhibit extremely low intrinsic photon decay rates ($\Gamma_0 = \omega_0/Q_0$) of $\Gamma_{0p} = 0.15$ GHz, $\Gamma_{0s} = 0.14$ GHz, and $\Gamma_{0i} = 0.16$ GHz, due to the high quality of the single-crystalline silicon and optimization of the fabrication process. We now see the dual role that the high-Q cavity assumes in our system. Here, the cavity-enhancement serves to greatly strengthen the efficiency and purity of the photon generation process and effectively modifies the Rayleigh scattering cross section \cite{Weiss95,Mazzei07}, such that scattering is highly preferential between counter-propagating modes. Hence, photon pairs are created within the cavity and experience a strong coherent coupling between the forward and backward modes, providing a convenient platform for demonstrating the proposed quantum interference phenomena. 

\begin{figure*}[ht]
\begin{center}
\includegraphics[scale = 1]{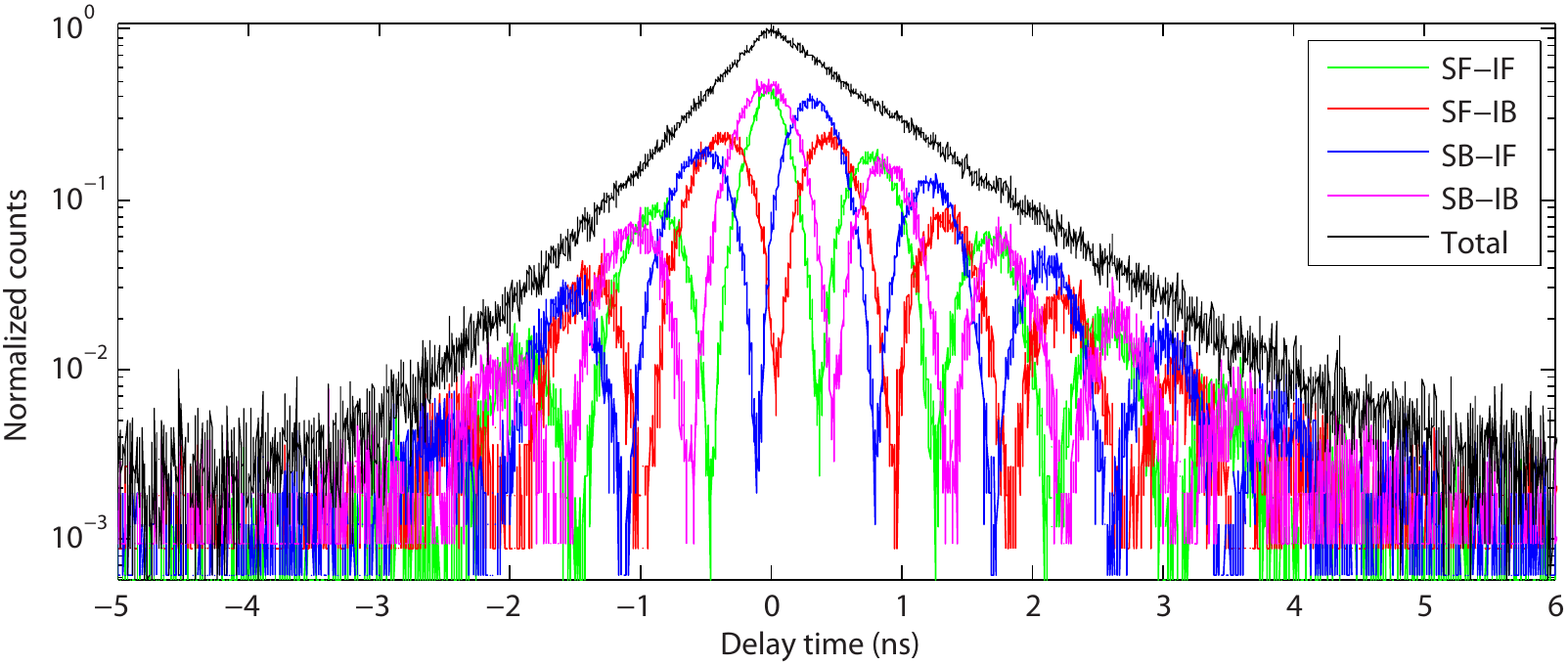}
\caption{ \small Time-evolving path-entanglement and high-contrast two-photon interference visibility. Here, we show the logarithmic scaling of pair correlations from all four path configurations superposed onto a single delay-time axis. The photon decay envelope, shown in black, results from the numeric sum of coincidence counts at each time bin. The strong quantum entanglement is observed through the coherent transfer of biphoton correlations between co-propagating (SF-IF (green) \& SB-IB (magenta)) and counter-propagating (SF-IB (red) \& SB-IF (blue)) states of the system.}
\label{Fig4}
\end{center}
\end{figure*}

\section*{Quantum interference and coherent oscillations}
Under the assumption of a closed system, the quantum states created within our device exhibit cyclically evolving path-entanglement (see Eq.~\ref{State}). We now consider what happens when the states are subjected to intrinsic loss, $\Gamma_{0m}$ (m = s,i), and are allowed to leave the microdisk at an external coupling rate, $\Gamma_{em}$.~Photon pairs are consequently transmitted into the two propagation directions of the optical waveguide (see Fig.~\ref{Fig1}(a)) and establish the following single photon pathways: signal forward (SF), signal backward (SB), idler forward (IF), and idler backward (IB). Moreover, the cyclically evolving path-entanglement within the device manifests as temporal correlations which coherently oscillate between pairs of propagation pathways, as seen in (see Appendix G),

\begingroup
\setlength{\abovedisplayskip}{3pt}
\setlength{\belowdisplayskip}{3pt}
\begin{align} \label{PairCorrelations}
p(t_{sj},t_{ik})  &= \hfill \nonumber \\ &N e^{-\Gamma_{tm}|\tau|}| \zeta_m^{jk} \cos(\beta_m\tau) + \eta_m^{jk} \sin(\beta_m\tau)|^2
\end{align}
\endgroup

\noindent where $t_{sj}$ (j = f,b) and $t_{ik}$ (k = f,b) respectively denote the times at which the signal and idler photons are emitted from the microdisk. The subscript $m = s$ when $t_{sj} > t_{ik}$, and $m = i$ when $t_{sj} < t_{ik}$. N is a constant quantifying properties of the device (see Appendix G), $\Gamma_{tm} = \Gamma_{0m} + \Gamma_{em}$ is the total photon decay rate, and  $\tau \equiv t_{sj} - t_{ik}$ denotes the delay between signal and idler emission times.  The strength of the oscillatory terms in Eq.~\ref{PairCorrelations} are respectively governed by $\zeta_m^{jk} = c_{m1}^{jk} |a_{pf}|^2 e^{-i\phi} - c_{m2}^{jk} |a_{pb}|^2$ and $\eta_m^{jk} = c_{m3}^{jk} |a_{pf}|^2 e^{-i\phi} - c_{m4}^{jk} |a_{pb}|^2$, with $|a_{pf}|^2$ and $|a_{pb}|^2$ being the energy contained in the forward and backward pump cavity modes, and $\phi$ defining their relative phase. The $c_{mn}^{jk}$ (n = 1 to 4) are constants defined by the coupling and decay rates of the modes (see Appendix G). The compact notation used in Eq.~\ref{PairCorrelations} emphasizes that although there are many intricate interactions occurring within the device, the pair correlations broadly consist of three main components, in that, they oscillate within a decay envelope set by the cavity photon lifetime, exhibit an oscillation frequency which matches the modal coupling rate, and are manipulated by interfering the counter-propagating intracavity pump waves.

\begin{figure*}[ht]
\begin{center}
\includegraphics{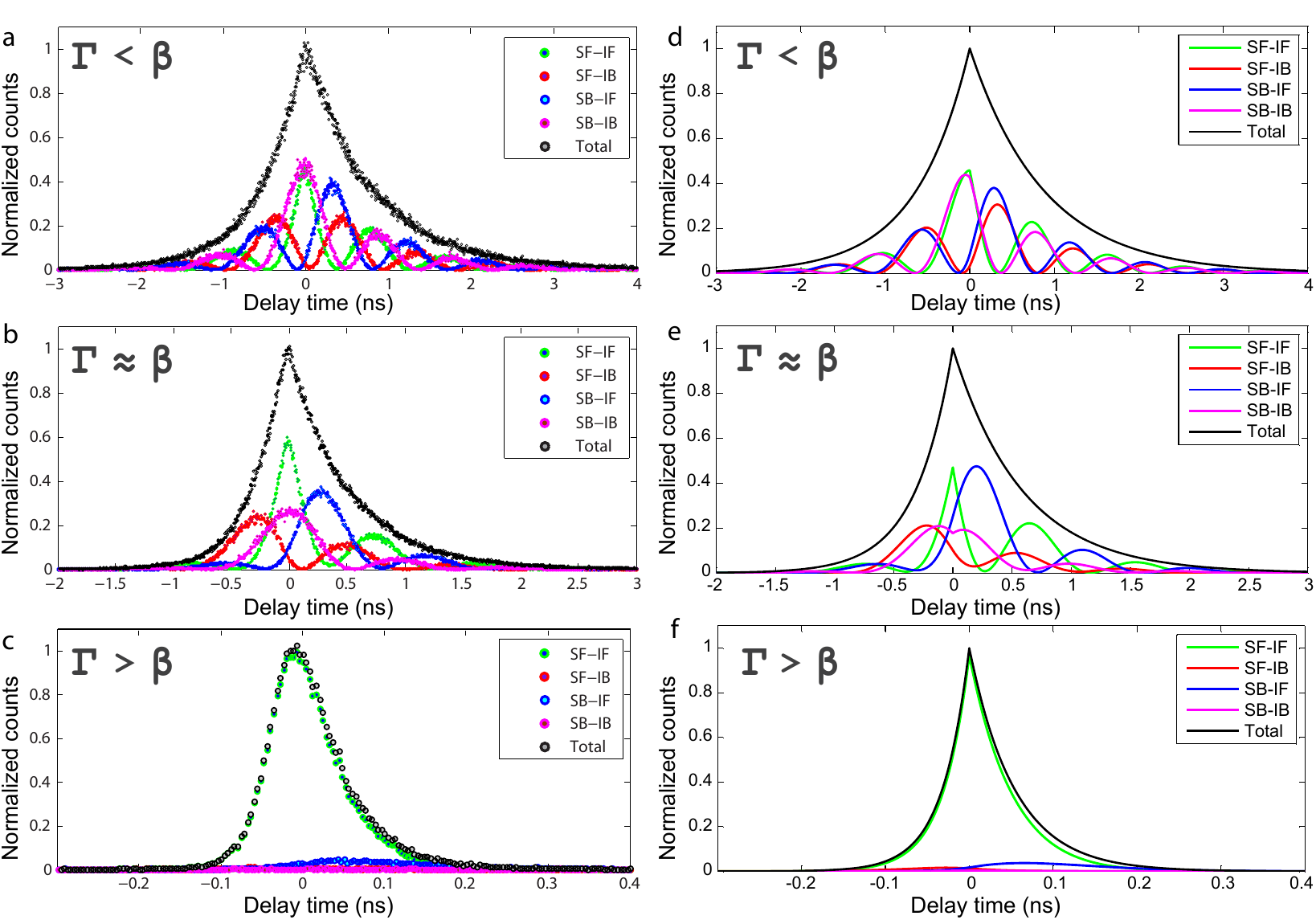}
\caption{ \small Manipulating the quantum state and entanglement by tuning the cavity photon lifetime. Measured pair correlations (without background subtraction) from each path configuration with the total photon decay set to be (a) less than the modal coupling rate, (b) similar to the modal coupling rate, and (c) greater than the modal coupling rate. (d) - (f) Theory plots corresponding to the experimental data in (a) - (c), respectively. In each figure, cross-correlation waveforms between signal forward - idler forward (SF-IF) are shown in green, signal forward - idler backward (SF-IB) are shown in red, signal backward - idler forward (SB-IF) are shown in blue, and signal backward - idler backward (SB-IB) are shown in magenta. The decay envelopes, shown in black, result from the numeric sum of coincidence counts from all path configurations at each time bin. All data are normalized to the peak of the photon decay envelope. We note that the biphoton coherence time in (c) is comparable to the detector response time, which causes the measured waveforms to display a Gaussian shape.}
\label{Fig5}
\end{center}
\end{figure*}

The total photon decay rate is conveniently tuned by varying the gap between the microdisk and waveguide. Initially, we set the total photon decay rate to be significantly less than the modal coupling rate and record the pair correlations for each of the four path configurations, as shown in Fig.~\ref{Fig3}. The measured correlations exhibit striking differences when compared to the monotonically decaying correlations between photon pairs from all other chip-scale sources studied to date. As predicted (see Eq~\ref{PairCorrelations}), the photon pairs in each of the path configurations exhibit coherent oscillations with estimated oscillation frequencies that are in good agreement with the measured doublet splittings. Furthermore, we see that biphotons in the co-propagating states (SF-IF \& SB-IB) are highly correlated at zero delay-time and then oscillate between being highly correlated and uncorrelated (Fig.~\ref{Fig3}(a),(d)). In contrast, a complementary effect is observed between biphotons in the counter-propagating states (SF-IB \& SB-IF)(Fig.~\ref{Fig3}(b),(c)). Taken together, it is apparent that as pair correlations are diminishing in one state they are intensifying in another, and vice versa. To precisely characterize this relationship, we superpose the correlations from each path configuration and numerically sum the coincidence counts at each time bin, resulting in the black correlation waveform (Fig.~\ref{Fig3}(e)). Here, we observe the remarkable effect that the pair correlations from all four states perfectly sum to give an exponential decay envelope, as predicted in Eq.~\ref{PairCorrelations}. 

\begin{figure*}[ht]
\begin{center}
\includegraphics[scale=0.95]{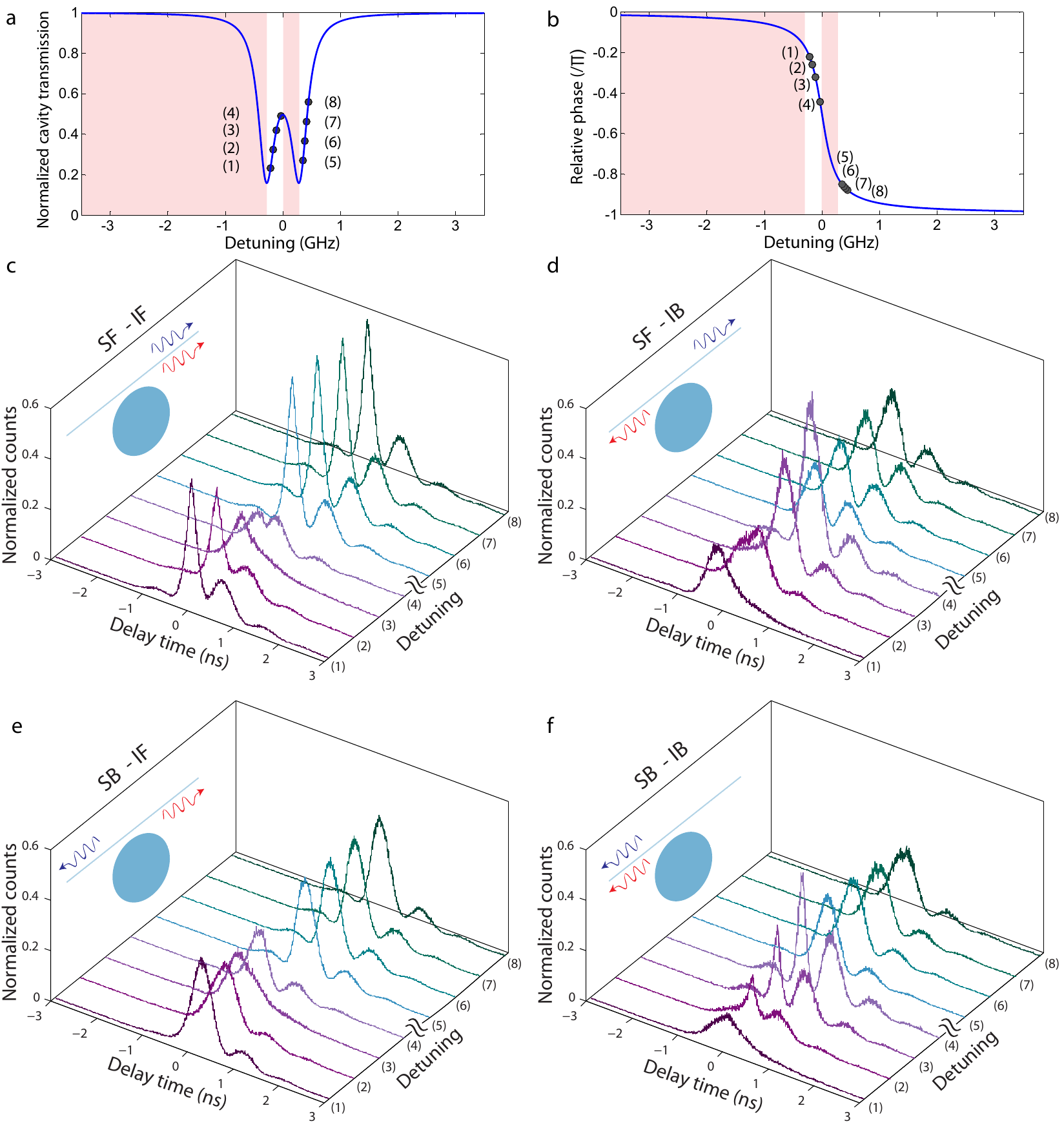}
\caption{ \small Tuning the intracavity quantum state through pump-induced quantum interference. Theoretical plots of the pump (a) transmission profile and (b) relative phase between counter-propagating modes. The external laser is locked to the pump cavity mode at points (1)-(8) which correspond to detuning values of $\Delta$ = (-0.21, -0.16, -0.11, -0.06, 0.35, 0.38, 0.41, 0.44) GHz, respectively, and denote the values appearing in (c)-(f). The laser locking is achieved through the thermo-optic feedback method where the light-red shading indicates the unstable locking regions. Cross-correlation waveforms (without background subtraction), measured between (c) signal forward - idler forward (SF-IF), (d) signal forward - idler backward (SF-IB), (e) signal backward - idler forward (SB-IF), and (f) signal backward - idler backward (SB-IB) are plotted as a function of detuning between the pump laser and resonance frequency of the pump cavity mode. Note that the waveforms within each stacked plot are offset by an equal spacing to aid in viewing clarity. Insets depict path configuration.}
\label{Fig6}
\end{center}
\end{figure*}

The logarithmic scaling in Fig.~\ref{Fig4} confirms the exponential nature of the photon decay envelope and reveals that the quantum interference responsible for driving the time-domain oscillations is of extremely high visibility. Here, we clearly observe that each of the correlation waveforms exhibit extinction ratios approaching or greater than 20 dB. Additionally, we see that the coincidence counts from the co-propagating states are almost completely out of phase with the counts from the counter-propagating states, suggesting that path-entanglement is present in the system. Recently, Du et al. showed that the arrival times between pairs of detection events may be used as a local parameter setting for entanglement measurements in systems where the photon coherence times are much longer than the timing resolution of the detectors \cite{Du17}. Owing to the high-Q nature of our microcavity, the photon coherence times here are on the order of nanoseconds, whereas the timing resolution of the superconducting detectors (see Appendix B) are only tens of picoseconds. Consequently, by varying delay-time as an analogue of phase, we record a Bell parameter (see Appendix F) of $S = 2.80 \pm 0.07$ which yields a Bell violation \cite{Bell64,Clauser69} of eleven standard deviations, confirming the path-entanglement predicted in Eq.\ref{State}. 


\section*{Modal coupling vs. photon decay rate}
A common feature among coherently coupled systems is the expression of markedly different behavior between the weak and strong coupling regimes \cite{Kimble92,Mabuchi02,Vahala03}. We explore these boundaries in Fig.~\ref{Fig5}, with the experimental data presented on the left half of the figure and the corresponding theory plots (see Appendix I) on the right.

By tuning the total photon decay rate, we significantly manipulate the intracavity quantum state, as is evidenced by the response of the pair correlations in Fig.~\ref{Fig5}. With the decay rate set to be less than the modal coupling rate, we experimentally observe the strong coupling regime, as shown in Fig.~\ref{Fig5}(a). Here, we see several high-contrast oscillations resulting from the strong quantum interference and path-entanglement. When the coupling and decay rates are similar, as is shown in Fig.~\ref{Fig5}(b), there are fewer oscillations, due to the diminished biphoton lifetime. Additionally, the pairs of correlation waveforms belonging to the same state classification - co-propagating vs.~counter-propagating - share less similarity than in Fig.~\ref{Fig5}(a). To explain this phenomenon, we note that changing the decay rate also changes the relationship between intracavity pump modes. In Fig.~\ref{Fig5}(a), the intracavity energies contained in the forward and backward pump modes are closer to being equalized, leading to a more symmetric system with respect to state generation, and consequently greater similarity between the pairs of waveforms within the same state classification. In the special case that the energies contained in the intracavity pump modes were exactly equal, then the pairs of correlation waveforms from the same state classification would be indistinguishable (see Appendix J). When the decay rate is set to be much greater than the coupling rate, we enter the weak coupling regime where the correlation waveforms exhibit pure monotonic decay, as shown in Fig.~\ref{Fig5}(c). In this extreme scenario, the quantum inference, path-entanglement and coherent oscillations are quenched as a result of tuning the biphoton state until it assumes the form of those produced in uncoupled systems.


\section*{Pump-induced quantum interference}
The interacting cavity modes each experience the same type of modal coupling. However, we emphasize that the coupling of pump modes uniquely affects the system, as they are the progenitors of the creation process. And as such, tuning properties of the coupled pump modes translates to directly manipulating the intracavity quantum state (see Eq.~\ref{State}) and the correlations that consequently manifest (see Eq.~\ref{PairCorrelations}). 

Changing the frequency of the external pump laser ($\omega_{L}$) relative to the resonance frequency of the pump cavity mode ($\omega_{0p}$) introduces a detuning, $\Delta = \omega_{L} - \omega_{0p}$, to the system. When the detuning is swept, the coupled pump modes display the well-established doublet transmission profile, as seen in Fig.~\ref{Fig6}(a). Importantly, varying the detuning induces a phase change between the intracavity pump modes in a manner analogous to a driven harmonic oscillator. Due to the high-Q nature of the cavity, a small amount of detuning about resonance enables a $\pi$ relative phase shift, as shown in Fig.~\ref{Fig6}(b). Thus, the coupled system naturally admits two regimes of operation, negative detuning (see points (1)-(4) in Fig.~\ref{Fig6}(a),(b)) and positive detuning (see points (5)-(8) in Fig.~\ref{Fig6}(a),(b)), with a significant phase change between them. We note that these changes are an internal response of the coupled system, and thus, are achieved without requiring a second external laser coupled to the counter-propagating mode. 

The effects described above and illustrated in Fig.~\ref{Fig6}(a),(b) are purely classical with regards to the pump mode alone. However, considering the system as a whole reveals the manner in which the pump-coupling induces quantum interference.  Without coupling, there is no backward-propagating pump, and setting the corresponding coefficients to zero in the quantum state (see Eq.~\ref{State}) and pair correlations (see Eq.~\ref{PairCorrelations}) removes the mechanism to directly change their form. In this case, the system still exhibits quantum interference, but it does so in a fixed configuration. When the pump modes are coupled, it enables coherent control over the strength of interference between creation pathways, adding a dynamic feature to the system. Thus, sweeping the detuning causes pump-induced quantum interference which significantly manipulates the intracavity quantum state. This effect clearly manifests in the variable response of the pair correlations shown in Fig.~\ref{Fig6}(c)-(f). In fact, the interference is so strong that changing from negative to positive detuning (point (4) to (5)) causes the SF-IB (see Fig.~\ref{Fig6}(d)) and SB-IF (see Fig.~\ref{Fig6}(e)) states to flip from exhibiting highly correlated to completely uncorrelated behavior near zero delay-time. Detailed versions of this transition, for each path configuration, may be seen in Appendix K.      


\section*{Conclusion}
In this work, we proposed and demonstrated a new set of quantum interference phenomena that result from the creation and coherent conversion of quantum states between the propagating modes of an optical microcavity. The resulting photonic quantum states are highly versatile and exhibit cyclically evolving path-entanglement. Importantly, we showed that the states and entanglement are greatly manipulated by controlling parameters of the device. We envision that the concepts presented here will have broad impacts relating to quantum state generation and novel entanglement properties, particularly in the field of quantum information processing. 

We note that the coupling of counter-propagating modes was achieved through Rayleigh scattering, which relies upon the existence of roughness at the surface of the device. Alternatively, the coupling could be established by bringing a nanoscale probe in contact with the device \cite{Yang10}, or through a technique called selective mode splitting (SMS), which couples counter-propagating modes by matching their azimuthal mode number with a periodic modulation along the perimeter of the microresonator \cite{Lu14}. These techniques would extend the concepts presented here to devices made from ultra-smooth materials and have the additional benefit of providing a means to control and tune the coupling rates.  

We showed that a single external laser can directly manipulate the quantum state through pump-induced quantum interference. Alternatively, the laser could be split and coupled into the device from both directions which would precisely control the amplitude and phase relationship between the internal pump modes independent of the cavity properties. This added feature would enable complete control over the quantum state (see Eq.~\ref{State}) and perfect pump-induced quantum interference visibility (see Appendix J). Additionally, this type of dramatic state manipulation would manifest in the coherent oscillations as a phenomenon akin to the Rabi flop operations that are used to optically prepare and control electron spins in solid-state systems \cite{Golter14}.   

Looking forward, a particularly intriguing follow-up involves building upon the capabilities of a single device by combining many devices in a large-scale photonic quantum circuit. Device to device and device to waveguide interactions could then be established and flexibly controlled through the addition of micro-electro-mechanical (MEMS) actuators, phase shifters, and micro-heaters, among other commonly used integrated photonics components. Based upon the observations made in this work, it is envisioned that such an architecture could be used to create exotic multi-photon states with fascinating properties, including controllable multi-partite entanglement and topologically protected quantum state transfer \cite{Soljacic14}.

\begin{acknowledgments}
The authors would like to thank John C. Howell and Sultan A. Wadood for helpful discussions. This work was supported by the National Science Foundation under Grant No.~ECCS-1351697 and EFMA-1641099.~It was performed in part at the Cornell NanoScale Facility, a member of the National Nanotechnology Coordinated Infrastructure (NNCI), which is supported by the National Science Foundation (Grant ECCS-1542081). 
\end{acknowledgments}


\section*{Appendices}
\appendix
\section{Device fabrication} 
The microdisk device is fabricated from a silicon-on-insulator wafer with a top silicon layer of 260 nm and buried oxide thickness of 2 $\mu$m. The initial device pattern is written into a high resolution electron sensitive resist (ZEP 520A) using electron-beam lithography. The pattern is then transferred to the silicon layer through an inductively-coupled-plasma (ICP) reactive-ion-etch (RIE), utilizing a SF$_6$/C$_4$F$_8$ gas chemistry. Finally, the buried oxide layer is removed by wet etching in hydrofluoric (HF) acid, yielding the suspended microdisk device seen in Fig.~\ref{Fig2}(a).

\section{Pair generation and photon statistics} 
A continuous-wave tunable pump laser (Santec TSL-550C) is transmitted through a course-wavelength-division-multiplexing (CWDM) multiplexer (MUX), having a 3-dB bandwidth of 17 nm and band isolation exceeding 120 dB, in order to prevent amplified spontaneous emission (ASE) from leaking into the single photon channels. The filtered pump light is then evanescently coupled from a tapered optical fiber into the device, using a nanopositioning setup. The polarization state at the point of coupling is controlled using fiber polarization controllers (FPC) in order to excite the quasi-TM pump cavity mode. The input optical power is set to $P_{in}$ = 8.54 $\mu$W for all experimental data presented here, and the optical power coupled into the cavity ranges from $P_{d,min}$ = 4.46 $\mu$W to $P_{d,max}$ = 6.44 $\mu$W, as a function of laser-cavity detuning. Photon pairs are coupled from the microdisk back into the tapered optical fiber (see Fig.~\ref{Fig1}(a)) in both the forward and backward directions and separated using CWDM demultiplexers. The demultiplexed pump beam is detected using a fast photodector, which allows for continuous monitoring of the coupled optical power, laser-cavity detuning, and implementation of the thermo-optic locking method \cite{Vahala04}. The signal and idler photons, from both propagation directions, are then passed through standard telecom optical switches in order to select the four path configurations based on the switch settings. The photons exiting the switches are passed through tunable bandpass filters (TBPF), having a 3-dB bandwidth of 1.2 nm, in order to suppress the Raman noise photons that are generated throughout the input side of the optical fiber. The single photons are then detected using two superconducting nanowire single photon detectors (SNSPD, SingleQuatum), which have extremely small timing jitters of 16 ps and detection efficiencies of 54\%. The photon arrival times are recorded using the time-tagged mode of a time-correlated single photon counter (TCSPC, PicoHarp 300) with time bins of 4 ps. The correlation waveforms appearing in Fig.~\ref{Fig3}, Fig.~\ref{Fig4}, and Fig.~\ref{Fig5}(a) were acquired over a period of $T_{acq}$ = 1800 seconds. The correlation waveforms appearing in Fig.~\ref{Fig5}(b),(c) and Fig.~\ref{Fig6} were acquired over a period $T_{acq}$ = 200 seconds. All coincidence counts presented in the Article are normalized to the peak of their respective decay envelope. A schematic of the experimental setup may be found in Appendix L.  \\

\section{The system Hamiltonian}
Here, we describe the Hamiltonian which governs the photon pair creation process between the coupled counter-propagating modes of an optical microresonator, as well as the coupling between the device and an optical waveguide. This Hamiltonian will serve as the springboard for our theoretical treatment of the intracavity quantum state (see Appendix D) and second-order correlations that manifest (see Appendix G). 

\begin{figure*}[ht!]
\begin{center}
\includegraphics[scale = 1.48]{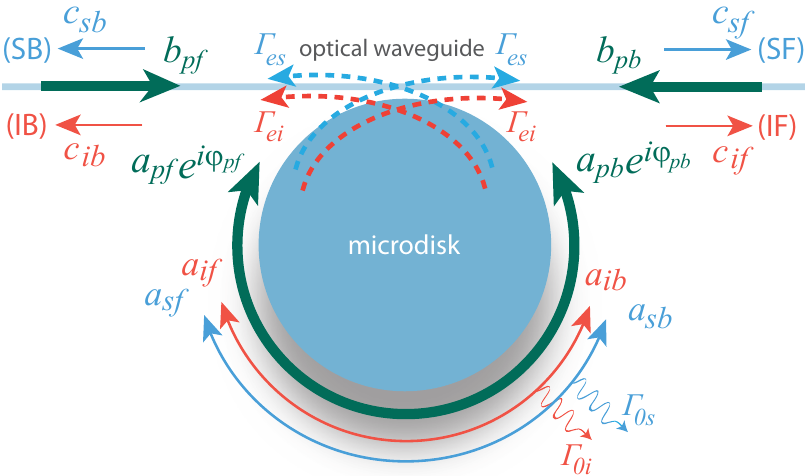}
\caption{ \small Photon pair creation between coherently coupled counter-propagating whispering-gallery modes (WGM). Forward, $b_{pf}$, and backward, $b_{pb}$, propagating pump waves (green) evanescently couple into the microdisk, on resonance with the pump cavity mode, respectively building up strong intracavity fields in the forward, $a_{pf}$, and backward, $a_{pb}$, directions. The counter-propagating intracavity pump fields are mutually coupled at a rate of $\beta_p$. Through spontaneous four-wave mixing (SFWM), signal (blue) and idler (red) photons are created within their respective cavity modes. Furthermore, the counter-propagating intracavity signal (idler) fields are mutually coupled at a rate of $\beta_s$ ($\beta_i$). Thus, the photon pairs coherently cycle between propagation directions within the cavity and ultimately exit into the four transmitted fields of the optical waveguide: $c_{sf}$ for signal forward (SF), $c_{sb}$ for signal backward (SB), $c_{if}$ for idler forward (IF), and $c_{ib}$ for idler backward (IB). Hence, correlations are established between photon pairs in the four path configurations: SF-IF, SF-IB, SB-IF, SB-IB.}
\label{physical_picture}
\end{center}
\end{figure*}

We begin by considering the establishment of Kerr-type nonlinear optical interactions between the pump (p), signal (s), and idler (i) whispering-gallery modes (WGMs), as illustrated in Fig.~\ref{physical_picture}. The passive cavity modes are specified by their resonance frequencies, $\omega_{\rm 0m}$ (m = p,s,i), along with their intrinsic decay rates, $\Gamma_{\rm 0m}$, external coupling rates, $\Gamma_{\rm em}$, and resonance splitting, $2 \beta_m$. Furthermore, input pump waves (at a carrier frequency of $\omega_{\rm p}$) propagating in the forward, $b_{\rm pf}$, and backward, $b_{\rm pb}$, directions are coupled from the optical waveguide into the device and coherently build up in their respective intracavity modes. In particular, we assume that there exists a mutual coupling between intracavity optical fields propagating in the forward (clockwise), $a_{\rm mf}$, and backward (counterclockwise), $a_{\rm mb}$, directions, which renormalizes the pairs of degenerate traveling-wave modes into pairs of standing-wave modes. The coupling occurs at a rate of $\beta_{\rm m}$, so that the standing-wave modes have distinct frequencies, $\omega_{0m}^{\pm} = \omega_{0m} \pm |\beta_m|$, which are shifted from the uncoupled system. When the intracavity pump waves undergo spontaneous four-wave mixing (SFWM), signal and idler photon pairs are created in their respective coupled counter-propagating modes. The Hamiltonian, $H = H_0 + H_I$, describing the relevant Kerr nonlinear interactions within the cavity is given by \cite{Jiang15}

\begin{widetext}
\begin{eqnarray}
H_0 = && \sum_{m=p,s,i}{ \left\{ \hbar \omega_{\rm 0m} \left(a_{\rm mf}^\dag a_{\rm mf}^{} + a_{\rm mb}^\dag a_{\rm mb}^{} \right) - \left( \hbar \beta_m a_{\rm mf}^\dag a_{\rm mb}^{} + \hbar \beta_m^* a_{\rm mb}^\dag a_{\rm mf}^{} \right) \right.}\nonumber \\
&& {\left. - \hbar \sqrt{\Gamma_{\rm em}} \left( (a_{\rm mf}^\dag b_{\rm mf}^{} + a_{\rm mb}^\dag b_{\rm mb}^{}) e^{-i\omega_m t} + (b_{\rm mf}^\dag a_{\rm mf}^{} + b_{\rm mb}^\dag a_{\rm mb}^{} ) e^{i\omega_m t} \right) \right\}}, \label{H0} \\
H_I =&& \frac{\hbar g_p}{2} \left( (a_{\rm pf}^\dag)^2 a_{\rm pf}^2 + (a_{\rm pb}^\dag)^2 a_{\rm pb}^2 + 4 a_{\rm pf}^\dag a_{\rm pf}^{} a_{\rm pb}^\dag a_{\rm pb}^{} \right) \nonumber\\
 && + 2\hbar (a_{\rm pf}^\dag a_{\rm pf}^{} + a_{\rm pb}^\dag a_{\rm pb}^{} ) \left( g_{\rm ps} (a_{\rm sf}^\dag a_{\rm sf}^{} + a_{\rm sb}^\dag a_{\rm sb}^{} ) + g_{\rm pi} (a_{\rm if}^\dag a_{\rm if}^{} + a_{\rm ib}^\dag a_{\rm ib}^{} )  \right) \nonumber\\
 && + \hbar g_{\rm psi} \left( a_{\rm sf}^\dag a_{\rm if}^\dag a_{\rm pf}^2 + a_{\rm sb}^\dag a_{\rm ib}^\dag a_{\rm pb}^2\right) + \hbar g_{\rm psi}^* \left( (a_{\rm pf}^\dag)^2 a_{\rm sf}^{} a_{\rm if}^{} + (a_{\rm pb}^\dag)^2 a_{\rm sb}^{} a_{\rm ib}^{} \right), \label{HI}
\end{eqnarray}
\end{widetext}

\noindent where the intracavity field operators are normalized such that $a_{\rm mj}^\dag a_{\rm mj}$ (m = p,s,i and j = f,b) represent the photon number operators of the system. The input pump fields are normalized such that $b_{\rm mj}^\dag b_{\rm mj}$ denote the input photon fluxes and satisfy the commutation relation $[b_{\rm mj} (t),b_{\rm m'j'}^\dag (t')] = \delta_{\rm mm'} \delta_{\rm jj'} \delta (t - t')$. We note that the number of photons contained in the pump modes greatly exceed the signal and idler modes, so that self-phase modulation (SPM) and cross-phase modulation (XPM) initiated from these modes may be neglected. The vacuum coupling rates for the pump-initiated self-phase modulation (SPM), cross-phase modulation, and SFWM are respectively denoted as $g_{\rm p}$, $g_{\rm pm}$ (m = s,i), and $g_{\rm psi}$. However, given the similarity in field profiles and frequencies between the interacting cavity modes, we approximate $g_{\rm p} \approx g_{\rm pm} \approx g_{\rm psi} \equiv g = \frac{c \eta n_{\rm 2} \hbar \omega_{p} \sqrt{\omega_{s} \omega_{i}}}{n_{s} n_{i} \bar{V}}$, where c is the speed of light in vacuum, $\eta$ is the spatial overlap fraction of the interacting cavity modes, $n_2 = \frac{3 \chi^{(3)}}{4 \varepsilon_0 c n_p^2}$ is the Kerr nonlinear coefficient \cite{Boyd08}, $n_s$ ($n_i$) is the index of refraction of the microresonator at the signal (idler) wavelength, and $\bar{V}$ is the effective mode volume.

\section{Quantum state from perturbation theory}

\par To determine the biphoton quantum state inside our optical microresonator, first-order time-dependent perturbation theory was applied to the closed system Hamiltonian (see Appendix C) in an undepleted pump regime. At the instant of pair creation (here defined as $t=0$), the state takes the form $\ket{\psi(t=0)}= {\rm c_f} \ket{f}_\text{s} \ket{f}_\text{i}+ {\rm c_b} \ket{b}_\text{s} \ket{b}_\text{i}$, where $f$ denotes the forward-propagating traveling-wave mode and $b$ denotes the backward-propagating traveling-wave mode. The subscripts $s$ and $i$ differentiate the signal and idler photons. Complex coefficients ${\rm c_f}$ and ${\rm c_b}$ are set by the relative amplitude and phase between the pump's forward and backward traveling-wave modes. 
\par The initial state can be intuited by viewing the nonlinear interaction Hamiltonian $H_\text{I}$ (see Eq.~\ref{H0} \& \ref{HI} for the complete Hamiltonian) in the basis of forward and backward propagating modes. The terms that contribute to the creation of signal and idler photons via spontaneous four-wave mixing (SFWM) are given by $H_\text{SFWM}=\hbar g_{\text{psi}}\left(a_\text{sf}^{\dagger} a_\text{if}^{\dagger} a_\text{pf}^2 + a_\text{sb}^{\dagger} a_\text{ib}^{\dagger} a_\text{pb}^2 \right)$ where $g_{\text{psi}}$ is the vacuum coupling rate for SFWM. These terms allow for photon pairs to be created in the co-propagating directions only, as is required to conserve angular momentum.
\par We implement time-dependent perturbation theory to verify the initial state. The linear portion of the closed-system Hamiltonian is
\begin{align}
H_0=\sum_{\text{m=p,s,i}} \hbar \omega_{0\text{m}} \left( a_\text{mf}^{\dagger} a_\text{mf}^{} + a_\text{mb}^{\dagger} a_\text{mb}^{} \right)- \nonumber \\ \hbar \left(\beta_\text{m}^{} a_\text{mf}^\dagger a_\text{mb}^{} + \beta_\text{m}^{\ast} a_\text{mb}^{\dagger} a_\text{mf}^{} \right). \label{H0_simp}
\end{align}
\noindent where $\omega_{0\text{m}}$ is the center frequency of the $m^{\text{th}}$ resonance and $\beta_{\text{m}}$ is the coupling between forward and backward modes of the $m^{\text{th}}$ resonance. 
\par Meanwhile, the perturbative Hamiltonian is $H_\text{SFWM}$, given that the other terms in $H_\text{I}$ have a negligible effect in the weakly pumped system considered here. To first order, time-dependent perturbation theory then gives an unnormalized state
\begin{figure*}[ht!]
\begin{center}
\includegraphics[scale = 1]{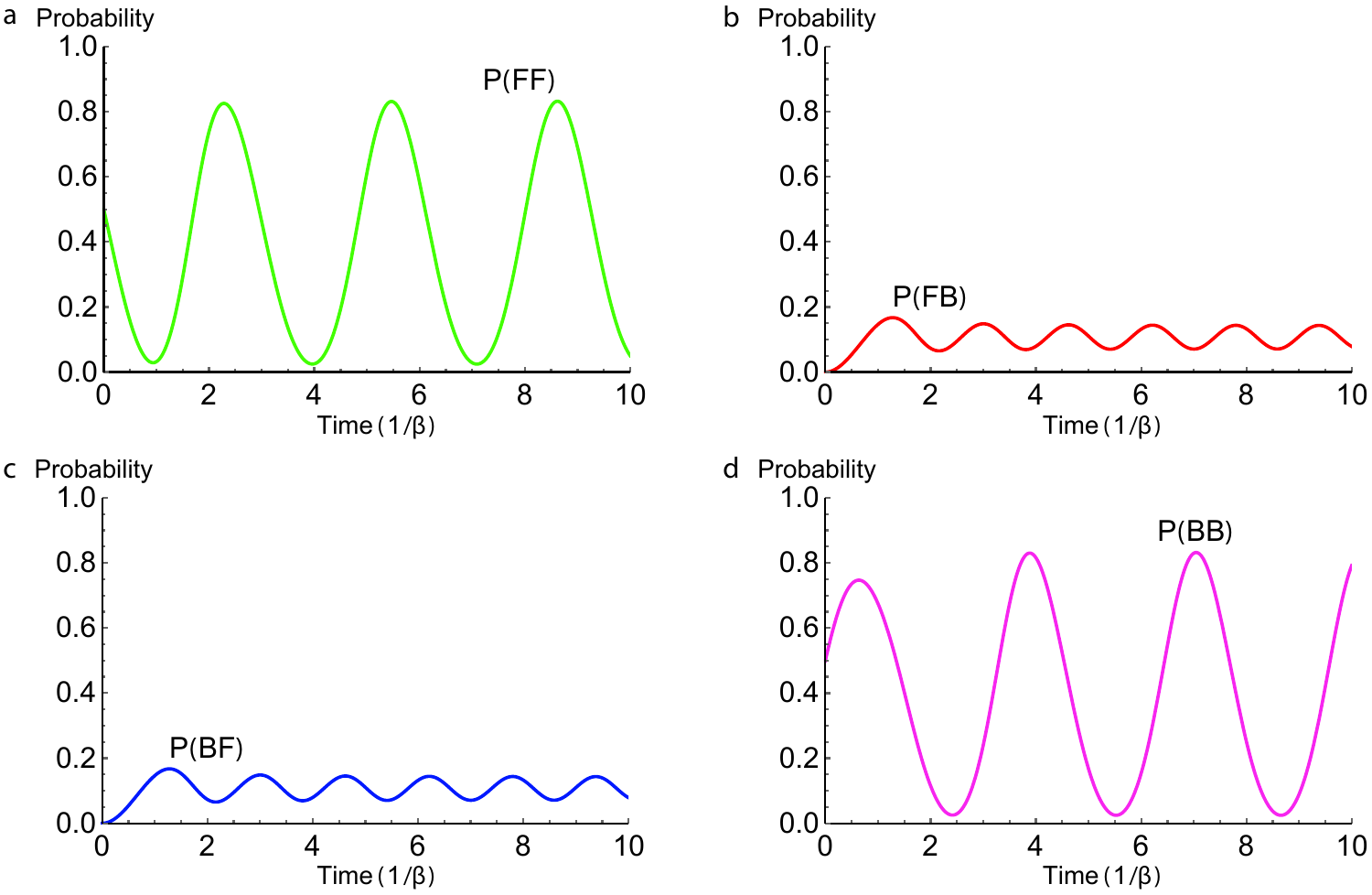}
\caption{A depiction of the time evolution of created signal and idlers assuming a closed system. To arrive at this result, $t=0$ was taken to be the instant of photon creation, forward and backward-propagating pumps were treated as equal amplitude, $\phi_\beta=\pi/4$, and $g_\text{psi}= 2 \pi \times 560$ radians per second. The probability of finding the photons in the path configurations (a) signal-forward and idler-forward, (b) signal-forward and idler-backward, (c) signal-backward and idler-forward, and (d) signal- backward and idler-backward is shown by the vertical axes. The horizontal axes display time in units of $1/\beta$ the reciprocal of the modal coupling. The state begins in a superposition of forward-forward and backward-backward path configurations, providing evidence for the initial state supplied in the main text.} \label{state_prob_plot}
\end{center}
\end{figure*}

\begin{equation} \label{psi_t}
\begin{split}
\ket{\psi (t)} &=  U_0(t,0) \ket{\text{Vac}} + \\ & \frac{1}{i \hbar}  \int_0^t \text{d}t^{\prime}  U_0(t,t^{\prime} ) H_{\text{SFWM}}(t^{\prime} ) U_0(t^{\prime} ,0) \ket{\text{Vac}}, 
\end{split}
\end{equation}

\noindent where $U_0(t_2, t_1)$ is defined to be the unitary time evolution operator $\text{exp}[-i H_0 (t_2-t_1)/\hbar]$. To evaluate the integral, it is helpful to diagonalize $H_0$ and write $H_\text{SFWM}$ in the basis that performs this diagonalization. In the microresonator system, this amounts to transforming into the standing-wave basis. The basis states, denoted by $+$ and $-$ are respectively the higher and lower energy eigenstates of $H_0$. Furthermore, we apply the approximation that $ \beta e^{i\phi_\beta} = \beta_\text{p}=\beta_\text{s}=\beta_\text{i} \equiv \beta$. Then the result of the transformation on $H_0$ and $H_\text{SFWM}$ is

\begin{widetext}
\begin{align}
&H_0 =  \sum_{\text{m=p,s,i}} \hbar \left( (\omega_{0\text{m}} - \beta ) a_{\text{m}-}^{\dagger} a_{\text{m}-}^{} + (\omega_{0\text{m}} + \beta  ) a_{\text{m}+}^{\dagger} a_{\text{m}+}^{} \right), \\
&H_{\text{SFWM}} = \frac{1}{2} \hbar g_\text{psi} \left( a_{\text{s}-}^{\dagger} a_{\text{i}-}^{\dagger} + a_{\text{s}+}^{\dagger} a_{\text{i}+}^{\dagger} e^{-2 i \phi_\beta} \right) \left( a_{\text{p}-}^2 + a_{\text{p}+}^2 e^{2 i \phi_\beta} \right) + \hbar g_\text{psi} \left( a_{\text{s}-}^{\dagger} a_{\text{i}+}^{\dagger}  + a_{\text{s}+}^{\dagger} a_{\text{i}-}^{\dagger} \right) a_{\text{p}-}^{} a_{\text{p}+}^{}.
\end{align}
\end{widetext}

Now Eq.~\ref{psi_t} becomes
\begin{align} \label{TimeDependentState}
\ket{\psi (t)} &= \ket{\text{Vac}} + \\ &\frac{1}{i \hbar}  \int_0^t \text{d}t ^{\prime}  e^{-i E_f (t-t^{\prime} )/\hbar} H_{\text{SFWM}}(t^{\prime} ) e^{-i E_i t^{\prime} } \ket{\text{Vac}}, \nonumber
\end{align}

\noindent where $E_i$ and $E_f$ are the energies of the initial and final states, respectively. Treating the pump as classical and undepleted, $E_i$ can be taken to zero by introducing a plane wave oscillation $\text{exp}[\mp 2 i \omega_p t']$ to $H_\text{SFWM}(t')$. $E_f$ is the total energy of the created signal and idler photons. The integral can then be evaluated, and will reflect a linear growth in time as the probability of photon creation increases. If we again define the instant of creation as $t=0$ and then normalize the first order terms, we can observe how the state of the created photons evolves in time. Figure \ref{state_prob_plot} shows the probability of detecting signal and idler photons in each of the four pairs of path configurations as a function of time.

\par
\begin{figure*}[ht]
\begin{center}
\includegraphics[scale = 0.498]{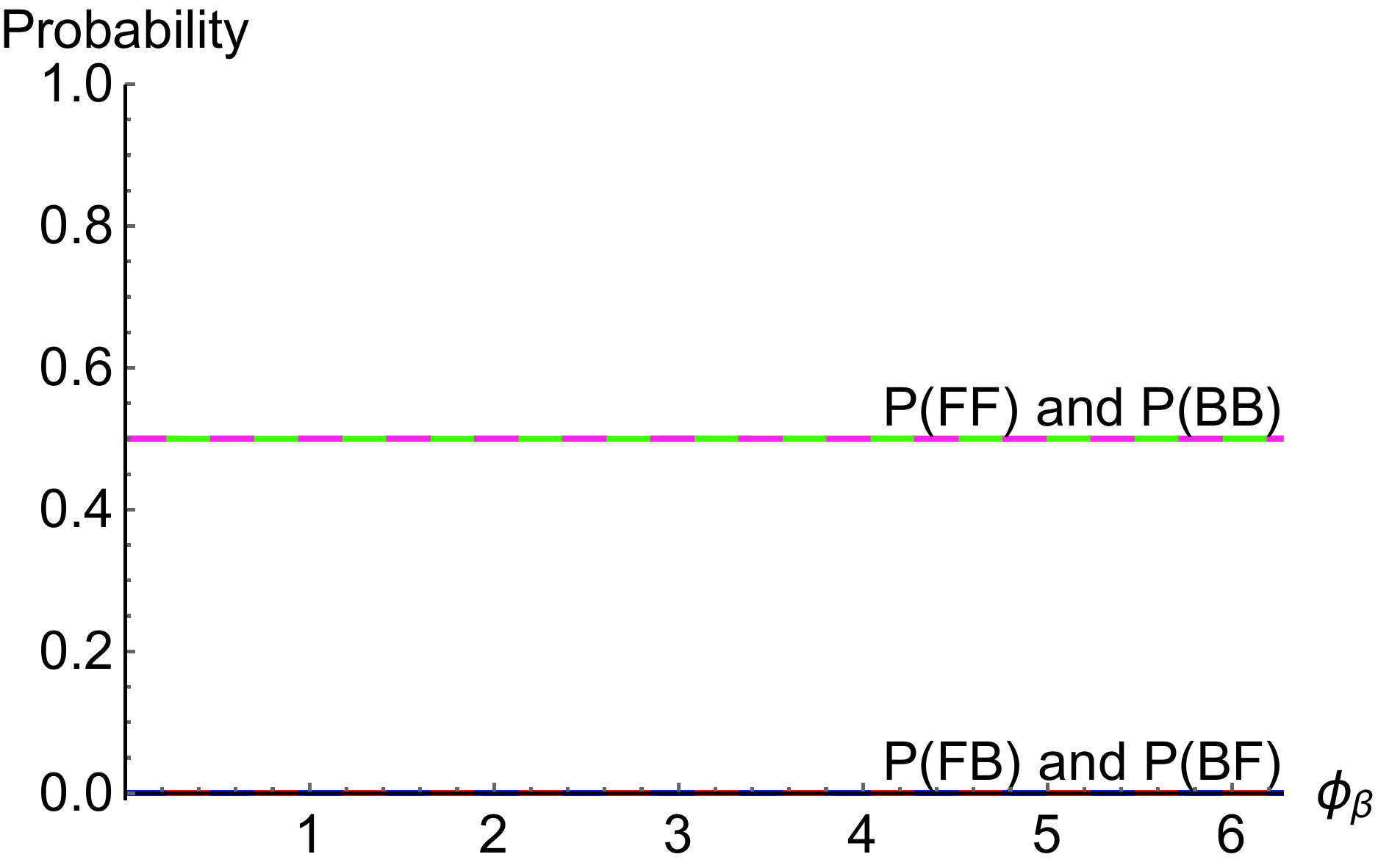}
\caption{Probability as a function of $\phi_\beta$ at $t=1/10000\beta$. The probability of created photons being in a co-propagating path configuration is $1/2$ for all values of $\phi_\beta$, while for counter-propagating path configurations it is 0 for all values of $\phi_\beta$.}
\label{phase_short_time}
\end{center}
\end{figure*}

 Because the probability of detecting counter-propagating signal and idler photon pairs is zero at $t=0$, we can affirm that the photons are created in a superposition of the forward-forward and backward-backward path configurations. That is, the initial biphoton state is indeed of the form $\ket{\psi(t=0)}= {\rm c_f} \ket{f}_\text{s} \ket{f}_\text{i}+ {\rm c_b} \ket{b}_\text{s} \ket{b}_\text{i}$, as stated in the main text. Notably, this is independent of $\phi_\beta$, as seen in Fig.~\ref{phase_short_time}. \\

\section{Quantum state evolution}
\par Beginning with the quantum state $\ket{\psi(t=0)}= {\rm c_f} \ket{f}_\text{s} \ket{f}_\text{i}+ {\rm c_b} \ket{b}_\text{s} \ket{b}_\text{i}$, time evolution can be easily performed by switching to the standing-wave basis using the following transformations,
\begin{align}
&\ket{f}_\text{m}=\frac{1}{\sqrt{2}}\left( \ket{-}_\text{m}-e^{i\phi_\beta}\ket{+}_\text{m} \right),\\
&\ket{b}_\text{m}=\frac{1}{\sqrt{2}}\left( e^{-i\phi_\beta}\ket{-}_\text{m} + \ket{+}_\text{m} \right).
\end{align}

\noindent The lower-energy standing wave mode is represented by $\ket{-}$. $\ket{+}$ likewise refers to the higher-energy standing-wave mode. For the following calculation, it will be assumed for simplicity that $\phi_\beta=0$. Implementing the basis transformation yields an initial state
\begin{widetext}
\begin{equation}
\ket{\psi(t=0)} =\frac{1}{2} \Big( \left( {\rm c_b} + {\rm c_f} \right) \big( \ket{-}_\text{s} \ket{-}_\text{i}+ \ket{+}_\text{s} \ket{+}_\text{i} \big) + \left({\rm c_b} - {\rm c_f} \right) \big( \ket{-}_\text{s} \ket{+}_\text{i} + \ket{+}_\text{s} \ket{-}_\text{i}  \big) \Big).
\end{equation}
\end{widetext}
\noindent Because the state is now in terms of energy eigenstates of the cavity, the unitary time evolution operator $U_0(t,0)=\text{exp}[-i H_0 t/\hbar]$ can be easily applied. The time evolved state in the standing-mode basis is thus,
\begin{widetext}
\begin{equation}
\ket{\psi(t)}=\frac{1}{2} e^{-i(\omega_s+\omega_i)t} \Big( \left( {\rm c_b} + {\rm c_f} \right) \big( e^{2 i \beta t } \ket{-}_\text{s} \ket{-}_\text{i}+  e^{-2 i \beta t } \ket{+}_\text{s} \ket{+}_\text{i} \big) + \left({\rm c_b} -{\rm c_f} \right) \big( \ket{-}_\text{s} \ket{+}_\text{i} + \ket{+}_\text{s} \ket{-}_\text{i}  \big) \Big).
\end{equation}
\end{widetext}
\noindent By transforming back to the basis of traveling waves and simplifying, this becomes

\begin{equation}
\begin{split}
\ket{\psi (t)} &=  e^{-i(\omega_s+\omega_i)t} \Big( \big( {\rm c_f} \cos^2 (\beta t) - {\rm c_b} \sin^2 (\beta t) \big) \ket{f}_\text{s} \ket{f}_\text{i} \\ &+ \big( {\rm c_b} \cos^2 (\beta t) - {\rm c_f} \sin^2 (\beta t) \big) \ket{b}_\text{s} \ket{b}_\text{i} + \\ & i ( {\rm c_f}+{\rm c_b} )\sin(2 \beta t) \big( \ket{f}_\text{s} \ket{b}_\text{i} + \ket{b}_\text{s} \ket{f}_\text{i}  \big) \Big). 
\end{split}
\end{equation}

\noindent This result appears as \textbf{Eq. 1} in the main text with the overall phase ignored.

\section{Bell test implementation}

Here, we present the methodology by which we computed a Bell parameter of $ S= 2.80 \pm 0.07$, violating the classical Clauser-Horne-Shimony-Holt (CHSH) inequality $\lvert S \rvert \leq 2$ \cite{Clauser69}. Typically, a Bell test entails varying two projective measurement angles, each between two different analyzer settings \cite{Bell64,Zeilinger12}. In the microresonator system, however, the test may be implemented by analogously varying the difference in arrival time between signal and idler photons $\tau=t_s-t_i$, as long as the photon coherence times are much larger than the temporal resolution of the detectors \cite{Du17}.
\par We generate a Bell correlation coefficient $E(\tau)$ in terms of second-order correlation functions $G^{(2)}_{jk}(\tau)$, where $j$ and $k$ denote the signal and idler path, respectively. Let f indicate the forward path and b the backward path. $E(\tau)$ is then 
\begin{equation}
E(\tau)=\frac{G^{(2)}_{\text{ff}}(\tau)+G^{(2)}_{\text{bb}}(\tau)-G^{(2)}_{\text{fb}}(\tau)-G^{(2)}_{\text{bf}}(\tau)}{G^{(2)}_{\text{ff}}(\tau)+G^{(2)}_{\text{bb}}(\tau)+G^{(2)}_{\text{fb}}(\tau)+G^{(2)}_{\text{bf}}(\tau)},
\end{equation} 

\noindent where $G^{(2)}_{jk}(\tau)$ is proportional to $p(t_{sj},t_{ik})$, given in Eq. 2 of the main text. This implies that the only time-dependence in $E(\tau)$ is in the form of sines and cosines of $\beta_m \tau$, where $\beta_m$ is the modal coupling, and $m$ refers to either the signal $s$ or idler $i$. We note that the exponential time-dependence in $p(t_{sj},t_{ik})$ cancels in $E (\tau)$.
\par Now we allow $\tau$ to vary as a function of two angles $\theta_a$ and $\theta_b$. This will enable us to vary $\tau$ in a manner equivalent to the usual Bell test angle variation. To mimic the angle-dependence of the typical Bell test, $\tau (\theta_a,\theta_b)$ must have the form \cite{Du17}
\begin{equation}
\tau (\theta_a, \theta_b)= \begin{cases} \frac{1}{\beta_i (\theta_a-\theta_b)} &\mbox{if } \theta_a-\theta_b < 0, \\ 
\frac{1}{\beta_s (\theta_a-\theta_b)} & \mbox{if } \theta_a-\theta_b \geq 0. \end{cases} \label{BellAngles}
\end{equation}
The Bell parameter can now be written as
\begin{align}
S &=  \lvert E (\tau (\theta_1, \theta_2 )) - E (\tau (\theta_1, \theta'_2 )) \rvert \hfill \nonumber \\ &+ \lvert E (\tau (\theta'_1, \theta_2 )) + E (\tau (\theta'_1, \theta'_2 )) \rvert.
\end{align}
Here $\theta_1$ and $\theta'_1$ are the measurement settings for $\theta_a$ while $\theta_2$ and $\theta'_2$ are the measurement settings for $\theta_b$. The settings are defined such that
\begin{equation}
\theta'_n=\theta_n -\pi/4,
\end{equation}
for $n=1,2$. We achieved our greatest violation of the CHSH inequality for $\theta_1=4.13$, $\theta'_1=3.34$, $\theta_2=0$, $\theta'_2=-0.79$. The resulting Bell parameter is $S=2.80$, well above the classical limit of $2$ but still within the quantum bound of $2\sqrt{2}$.
\par To determine error in our Bell parameter, we simulated 2500 data sets under the assumption of Poissonian counting statistics, using the number of real counts in each 4 ps time bin as mean. For each simulated data set, the Bell parameter producing the largest CHSH violation was computed. The standard deviation of these simulated Bell parameters was $0.07$, giving the result $S=2.80 \pm 0.07$ that appears in the main text. We note that the values of $\beta_m$ (m = s,i), appearing in Eq.~\ref{BellAngles}, were measured using a calibrated Mach-Zehnder interferometer (MZI). The uncertainty associated with these measurements was so small in comparison to the photon counting uncertainty that it had no bearing on the overall error of the Bell parameter.


\section{Theory of coherent oscillations in photon pair correlations}

In Appendix E, we showed that the quantum states generated within the microresonator exhibit cyclically evolving path-entanglement (see Eq.~\ref{TimeDependentState}). When the photon pairs exit the system (see Fig.~\ref{physical_picture}), they are transmitted into both directions of the optical waveguide which establishes the following single photon pathways: signal forward (SF), signal backward (SB), idler forward (IB), and idler backward (IB). The cyclically evolving path-entanglement within the microresonator will consequently manifest as coherent oscillations between pairs of propagation pathways. To theoretically demonstrate this phenomenon, we employ the Hamiltonian in Eq.~\ref{H0}-\ref{HI}, in order to write the Heisenberg-Langevin equations of motion. These equation describe the wave dynamics within the microdisk, and are given as 

\begin{widetext}
\begin{eqnarray}
\frac{da_{\rm pf}}{dt} &=& (-i\omega_{\rm 0p} - \Gamma_{\rm tp}/2) a_{\rm pf}^{} + i \beta_p a_{\rm pb}^{} - i g (a_{\rm pf}^\dag a_{\rm pf}^{} + 2 a_{\rm pb}^\dag a_{\rm pb}^{}) a_{\rm pf}^{} + \zeta_{\rm pf}(t), \label{dapfdt} \\
\frac{da_{\rm pb}}{dt} &=& (-i\omega_{\rm 0p} - \Gamma_{\rm tp}/2) a_{\rm pb}^{} + i \beta_p^* a_{\rm pf}^{} - i g (a_{\rm pb}^\dag a_{\rm pb}^{} + 2 a_{\rm pf}^\dag a_{\rm pf}^{}) a_{\rm pb}^{} + \zeta_{\rm pb}(t), \label{dapbdt} \\
\frac{da_{\rm sf}}{dt} &=& (-i\omega_{\rm 0s} - \Gamma_{\rm ts}/2) a_{\rm sf}^{} + i \beta_s a_{\rm sb}^{} - 2 i g (a_{\rm pf}^\dag a_{\rm pf}^{} + a_{\rm pb}^\dag a_{\rm pb}^{} ) a_{\rm sf}^{} - i g a_{\rm if}^\dag a_{\rm pf}^2 + \zeta_{\rm sf}(t), \label{dasfdt} \\
\frac{da_{\rm sb}}{dt} &=& (-i\omega_{\rm 0s} - \Gamma_{\rm ts}/2) a_{\rm sb}^{} + i \beta_s^* a_{\rm sf}^{} - 2 i g (a_{\rm pf}^\dag a_{\rm pf}^{} + a_{\rm pb}^\dag a_{\rm pb}^{} ) a_{\rm sb}^{} - i g a_{\rm ib}^\dag a_{\rm pb}^2 + \zeta_{\rm sb}(t), \label{dasbdt} \\
\frac{da_{\rm if}}{dt} &=& (-i\omega_{\rm 0i} - \Gamma_{\rm ti}/2) a_{\rm if}^{} + i \beta_i a_{\rm ib}^{} - 2 i g (a_{\rm pf}^\dag a_{\rm pf}^{} + a_{\rm pb}^\dag a_{\rm pb}^{} ) a_{\rm if}^{} - i g a_{\rm sf}^\dag a_{\rm pf}^2 + \zeta_{\rm if}(t), \label{daifdt} \\
\frac{da_{\rm ib}}{dt} &=& (-i\omega_{\rm 0i} - \Gamma_{\rm ti}/2) a_{\rm ib}^{} + i \beta_i^* a_{\rm if}^{} - 2 i g (a_{\rm pf}^\dag a_{\rm pf}^{} + a_{\rm pb}^\dag a_{\rm pb}^{} ) a_{\rm ib}^{} - i g a_{\rm sb}^\dag a_{\rm pb}^2 + \zeta_{\rm ib}(t), \label{daibdt}
\end{eqnarray} 
\end{widetext}

\noindent where $\zeta_{\rm mj}(t) \equiv i \sqrt{\Gamma_{\rm em}} b_{\rm mj}(t) + \sqrt{\Gamma_{\rm 0m}} u_{\rm mj}(t)$ (m = p,s,i and j = f,b) and $u_{\rm mj}$ are noise operators associated with intrinsic cavity losses and obey the commutation relation $[u_{\rm mj} (t),u_{\rm m'j'}^\dag (t')] = \delta_{\rm mm'} \delta_{\rm jj'} \delta (t - t')$. 

In order to solve Eq.~\ref{dapfdt}-\ref{daibdt}, we treat the pumps as classical and undepleted (with respect to the Kerr nonlinear interactions), transform to a frame rotating at the carrier frequency of the input pump waves, and apply exponential transformations to eliminate the SPM and XPM terms, which gives

\begin{widetext}
\begin{eqnarray}
\frac{da_{\rm pf}}{dt} &=& (i\Delta - \Gamma_{\rm tp}/2) a_{\rm pf}^{} + i \beta_p a_{\rm pb}^{} + i \sqrt{\Gamma_{\rm ep}} b_{\rm pf}^{}, \label{dapfdt_2} \\
\frac{da_{\rm pb}}{dt} &=& (i\Delta - \Gamma_{\rm tp}/2) a_{\rm pb}^{} + i \beta_p^* a_{\rm pf}^{} + i \sqrt{\Gamma_{\rm ep}} b_{\rm pb}^{}, \label{dapbdt_2} \\
\frac{da_{\rm sf}}{dt} &=& (i\Delta - \Gamma_{\rm ts}/2) a_{\rm sf}^{} + i \beta_s a_{\rm sb}^{}  - i g a_{\rm if}^\dag a_{\rm pf}^2 + \zeta_{\rm sf}(t), \label{dasfdt_2} \\
\frac{da_{\rm sb}}{dt} &=& (i\Delta - \Gamma_{\rm ts}/2) a_{\rm sb}^{} + i \beta_s^* a_{\rm sf}^{}  - i g a_{\rm ib}^\dag a_{\rm pb}^2 + \zeta_{\rm sb}(t), \label{dasbdt_2} \\
\frac{da_{\rm if}}{dt} &=& (i\Delta - \Gamma_{\rm ti}/2) a_{\rm if}^{} + i \beta_i a_{\rm ib}^{} - i g a_{\rm sf}^\dag a_{\rm pf}^2 + \zeta_{\rm if}(t), \label{daifdt_2} \\
\frac{da_{\rm ib}}{dt} &=& (i\Delta - \Gamma_{\rm ti}/2) a_{\rm ib}^{} + i \beta_i^* a_{\rm if}^{} - i g a_{\rm sb}^\dag a_{\rm pb}^2 + \zeta_{\rm ib}(t), \label{daibdt_2}
\end{eqnarray} 
\end{widetext}

\noindent where $\Delta \equiv \omega_{\rm p} - \omega_{\rm 0p}$ is the detuning between the carrier frequency of the input pump waves and the resonance frequency of the pump cavity mode. The coupled pump equations (Eq.~\ref{dapfdt_2} \& \ref{dapbdt_2}) may now be trivially solved due to the removal of any explicit time dependence. The remaining equations (Eq.~\ref{dasfdt_2}-\ref{daibdt_2}) are conveniently solved in the frequency domain by applying Fourier transforms, yielding

\begin{widetext}
\begin{eqnarray}
\tilde{\zeta}_{\rm sf}(\omega) &=& -\lbrace i(\omega + \Delta) - \Gamma_{\rm ts}/2 \rbrace \tilde{a}_{\rm sf}^{}(\omega) - i \beta_{\rm s} \tilde{a}_{\rm sb}^{}(\omega) + i g a_{\rm pf}^2 \tilde{a}_{\rm if}^\dag(- \omega),  \label{sfw} \\
\tilde{\zeta}_{\rm sb}(\omega) &=& -\lbrace i(\omega + \Delta) - \Gamma_{\rm ts}/2 \rbrace \tilde{a}_{\rm sb}^{}(\omega) - i \beta_{\rm s}^* \tilde{a}_{\rm sf}^{}(\omega) + i g a_{\rm pb}^2 \tilde{a}_{\rm ib}^\dag(- \omega),  \label{sbw} \\
\tilde{\zeta}_{\rm if}^\dag(- \omega) &=& -\lbrace i(\omega + \Delta) - \Gamma_{\rm ti}/2 \rbrace \tilde{a}_{\rm if}^\dag(- \omega) + i \beta_{\rm i}^* \tilde{a}_{\rm ib}^\dag(- \omega) - i g (a_{\rm pf}^*)^2 \tilde{a}_{\rm sf}^{}(\omega),  \label{ifw} \\
\tilde{\zeta}_{\rm ib}^\dag(- \omega) &=& -\lbrace i(\omega + \Delta) - \Gamma_{\rm ti}/2 \rbrace \tilde{a}_{\rm ib}^\dag(- \omega) + i \beta_{\rm i} \tilde{a}_{\rm if}^\dag(- \omega) - i g (a_{\rm pb}^*)^2 \tilde{a}_{\rm sb}^{}(\omega),  \label{ibw} 
\end{eqnarray} 
\end{widetext}

\noindent where $\Gamma_{\rm tm} = \Gamma_{\rm 0m} + \Gamma_{\rm em}$ (m = s,i) are the total photon decay rates for the cavity modes. Eq.~\ref{sfw}-\ref{ibw} may now be written using matrix formalism as, $\vec{\zeta}(\omega) = -\textbf{M} \vec{a}(\omega)$, with the following vector (matrix) components (elements)

\begin{widetext}
\begin{equation}
\begin{pmatrix}
\zeta_{\rm sf}(\omega) \\ \zeta_{\rm sb}(\omega) \\ \zeta_{\rm if}^\dag(- \omega) \\ \zeta_{\rm ib}^\dag(- \omega)
\end{pmatrix}
 = - 
 \begin{pmatrix}
  i(\omega + \Delta) - \Gamma_{\rm ts}/2 & i \beta_s e^{i \phi_{\beta_s}} & -i g a_{\rm pf}^2 & 0 \\
  i \beta_s e^{-i \phi_{\beta_s}} & i(\omega + \Delta) - \Gamma_{\rm ts}/2 & 0 & -i g a_{\rm pb}^2 \\
  i g^* (a_{\rm pf}^*)^2  & 0  & i(\omega + \Delta) - \Gamma_{\rm ti}/2 & - i \beta_i e^{-i \phi_{\beta_i}}  \\
  0 & i g^* (a_{\rm ab}^*)^2 & - i \beta_i e^{i \phi_{\beta_i}} & i(\omega + \Delta) - \Gamma_{\rm ti}/2
 \end{pmatrix} 
\begin{pmatrix}
a_{\rm sf}(\omega) \\ a_{\rm sb}(\omega) \\ a_{\rm if}^\dag(- \omega) \\ a_{\rm ib}^\dag(- \omega)
\end{pmatrix},  \label{M_Matrix}
\end{equation}
\end{widetext}

\noindent where we have dropped the tilde notation. Additionally, we have written the coupling terms as $\beta_m = \beta_m e^{i \phi_{\beta m}}$ to explicitly display their spatial phase dependence which results from the orientation of the standing-wave mode patterns to the point of coupling at the waveguide. Now, we can solve for the intracavity fields, noting that $\vec{a}(\omega) = \textbf{T} \vec{\zeta}(\omega)$, where $\textbf{T} \equiv -\textbf{M}^{-1}$. Finally, we relate the transmitted fields, $c_{\rm mj}$, to the input and intracavity fields through the standard input-output relations \cite{Walls08}, $c_{\rm mj} = b_{\rm mj} + i\sqrt{\Gamma_{\rm em}} a_{\rm mj}$.

The device presented in Fig.~\ref{physical_picture} couples photon pairs generated in the forward and backward directions, resulting in correlations between the following four configurations: signal forward - idler forward (SF-IF), signal forward - idler backward (SF-IB), signal backward - idler forward (SB-IF), and signal backward - idler backward (SB-IB). The correlations may be theoretically described using second-order temporal correlation functions \cite{Mandel95}, $p_{c,jk}(t_{\rm sj},t_{ik})$, with

\begin{equation} \label{paircorr_ift}
\begin{split}
p_{c,jk}(t_{\rm sj},t_{\rm ik}) &\equiv  \braket{c_{\rm ik}^\dag(t) c_{\rm sj}^\dag(t+\tau_{\rm jk}) c_{\rm sj}(t+\tau_{\rm jk}) c_{\rm ik}(t)} \\ &- \braket{c_{\rm sj}^\dag(t)c_{\rm sj}(t)} \braket{c_{\rm ik}^\dag(t)c_{\rm ik}(t)} \\ &= \Gamma_{\rm es} \Gamma_{\rm ei} |\frac{1}{2\pi} \int d\omega K_{\rm jk}(\omega) e^{-i \omega \tau_{jk}}|^2, 
\end{split}
\end{equation}


\noindent where $\tau_{\rm jk} \equiv t_{\rm sj} - t_{\rm ik}$ (j = f,b and k = f,b) denotes the delay time between the emission of signals and idlers from the microresonator into the waveguide (see Fig.~\ref{physical_picture}), and $K_{\rm jk}(\omega)$ are Kernel functions which depend upon the path configuration.  The Kernel functions, associated with the four path configurations, are defined as

\begin{widetext}
\begin{eqnarray}
K_{\rm ff} \equiv T_{31}^*(\omega)[\Gamma_{\rm ts} T_{11}(\omega) - 1] + \Gamma_{\rm ts} T_{12}(\omega)T_{32}^*(\omega), \label{K_ff} \\ 
K_{\rm fb} \equiv T_{41}^*(\omega)[\Gamma_{\rm ts} T_{11}(\omega) - 1] + \Gamma_{\rm ts} T_{12}(\omega)T_{42}^*(\omega), \label{K_fb} \\ 
K_{\rm bf} \equiv T_{32}^*(\omega)[\Gamma_{\rm ts} T_{22}(\omega) - 1] + \Gamma_{\rm ts} T_{21}(\omega)T_{31}^*(\omega), \label{K_bf} \\ 
K_{\rm bb} \equiv T_{42}^*(\omega)[\Gamma_{\rm ts} T_{22}(\omega) - 1] + \Gamma_{\rm ts} T_{21}(\omega)T_{41}^*(\omega), \label{K_bb}
\end{eqnarray}   
\end{widetext}

\noindent where $T_{\rm nl}$ are elements of the matrix \textbf{T}. After computing the matrix \textbf{T} and performing the inverse Fourier transform in Eq.~\ref{paircorr_ift} (both of these steps are accomplished in Mathematica), we arrive at the pair correlation functions for the four path configurations

\begin{widetext}
\begin{eqnarray} 
p_{c,ff}(\tau) = 
	\begin{cases}
	  N e^{\Gamma_{\rm ti} \tau} |\lbrace c_0 f e^{-i \phi} - c_1 b \rbrace \cos{(\beta_i \tau)} + \lbrace c_2 f e^{-i \phi} + c_3 b \rbrace \sin{(\beta_i \tau)}|^2 & \quad (\tau < 0), \\
    N e^{-\Gamma_{\rm ts} \tau} |\lbrace c_0 f e^{-i \phi} - c_1 b \rbrace \cos{(\beta_s \tau)} - \lbrace c_3 f e^{-i \phi} + c_2 b \rbrace \sin{(\beta_s \tau)}|^2 & \quad (\tau \ge 0), \label{paircorr_ff}
    \end{cases} \\ \nonumber \\
p_{c,fb}(\tau) = 
	\begin{cases}
	  N e^{\Gamma_{\rm ti} \tau} |\lbrace c_2 f e^{-i \phi} + c_3 b \rbrace \cos{(\beta_i \tau)} - \lbrace c_0 f e^{-i \phi} - c_1 b \rbrace \sin{(\beta_i \tau)}|^2 & \quad (\tau < 0), \\
    N e^{-\Gamma_{\rm ts} \tau} |\lbrace c_2 f e^{-i \phi} + c_3 b \rbrace \cos{(\beta_s \tau)} - \lbrace c_1 f e^{-i \phi} - c_0 b \rbrace \sin{(\beta_s \tau)}|^2 & \quad (\tau \ge 0), \label{paircorr_fb}
    \end{cases} \\ \nonumber \\
p_{c,bf}(\tau) = 
	\begin{cases}
	  N e^{\Gamma_{\rm ti} \tau} |\lbrace c_3 f e^{-i \phi} + c_2 b \rbrace \cos{(\beta_i \tau)} + \lbrace c_1 f e^{-i \phi} - c_0 b \rbrace \sin{(\beta_i \tau)}|^2 & \quad (\tau < 0), \\
    N e^{-\Gamma_{\rm ts} \tau} |\lbrace c_3 f e^{-i \phi} + c_2 b \rbrace \cos{(\beta_s \tau)} + \lbrace c_0 f e^{-i \phi} - c_1 b \rbrace \sin{(\beta_s \tau)}|^2 & \quad (\tau \ge 0), \label{paircorr_bf}
    \end{cases} \\ \nonumber \\
p_{c,bb}(\tau) = 
	\begin{cases}
	  N e^{\Gamma_{\rm ti} \tau} |\lbrace -c_1 f e^{-i \phi} + c_0 b \rbrace \cos{(\beta_i \tau)} + \lbrace c_3 f e^{-i \phi} + c_2 b \rbrace \sin{(\beta_i \tau)}|^2 & \quad (\tau < 0), \\
    N e^{-\Gamma_{\rm ts} \tau} |\lbrace -c_1 f e^{-i \phi} + c_0 b \rbrace \cos{(\beta_s \tau)} - \lbrace c_2 f e^{-i \phi} + c_3 b \rbrace \sin{(\beta_s \tau)}|^2 & \quad (\tau \ge 0), \label{paircorr_bb}
    \end{cases} 
\end{eqnarray}
\end{widetext}

\noindent where $f \equiv |a_{pf}(\Delta)|^2$ and $b \equiv |a_{pb}(\Delta)|^2$ are the optical energies contained in the forward and backward pump cavity modes, respectively, and $\phi \equiv 2 \phi_p + \phi_{\beta}$ defines the phase relationship between them. In particular, $\phi_p \equiv \phi_{pf} - \phi_{pb}$ and $\phi_{\beta} \equiv 2\phi_{\rm \beta p} - \phi_{\rm \beta s} - \phi_{\rm \beta i}$. We note that in the process of reaching these solutions we have made use of the fact that our system operates in the weak interaction limit ($gf \ll \Gamma_{\rm tp}$ and $gb \ll \Gamma_{\rm tp}$), such that terms beyond first order in $f$ and $b$ may be dropped. In Eq.~\ref{paircorr_ff}-\ref{paircorr_bb}, the constants $c_{\rm n}$ (n = 1 to 4) and N depend upon properties of the device in the following manner

\begin{eqnarray}
N &=& \frac{4 \Gamma_{es} \Gamma_{ei} \Gamma_t^4 g^2}{((c_0 - c_1)(c_0 + c_1))^2}, \label{N} \\ c_0 &=& (4 \beta_s^2 + 4 \beta_i^2 + \Gamma_t^2)\Gamma_t, \label{c0} \\ c_1 &=& 8 \beta_s \beta_i \Gamma_t, \label{c1} \\ c_2 &=& 8 \beta_i^3 - 8 \beta_s^2 \beta_i + 2 \beta_i \Gamma_t^2, \label{c2} \\ c_3 &=& 8 \beta_s^3 - 8 \beta_s \beta_i^2 + 2 \beta_s \Gamma_t^2, \label{c3}
\end{eqnarray}

\noindent where $\Gamma_t \equiv \Gamma_{ts} + \Gamma_{ti}$. Equations \ref{paircorr_ff}-\ref{paircorr_bb}, along with the constants defined in Eq.~\ref{N}-\ref{c3}, formulate the expanded version of \textbf{Eq. 2} that appears in the main text. 

\section{Obtaining the detuning-dependent intracavity pump fields}
As is evident in Eq.~\ref{paircorr_ff}-\ref{paircorr_bb}, the intracavity pump fields play a crucial role in controlling the structure of the pair correlations. Equations \ref{dapfdt_2} \& \ref{dapbdt_2} describe the coupled intracavity pump modes, and after assuming a steady state solution, may be written as,

\begin{eqnarray}
0 &=& (i\Delta - \Gamma_{\rm tp}/2) a_{\rm pf} + i \beta_p a_{\rm pb} + i \sqrt{\Gamma_{\rm ep}} b_{\rm pf}, \label{apf_ss} \\
0 &=& (i\Delta - \Gamma_{\rm tp}/2) a_{\rm pb} + i \beta_p^* a_{\rm pf} + i \sqrt{\Gamma_{\rm ep}} b_{\rm pb}, \label{apb_ss} 
\end{eqnarray}

\noindent or alternatively in the following matrix form,

\begin{equation}
\begin{pmatrix}
i \Delta - \Gamma_{\rm tp}/2 & i \beta_p e^{i \phi_{\beta_p}} \\ i \beta_p e^{-i \phi_{\beta_p}} & i \Delta - \Gamma_{\rm tp}/2
\end{pmatrix}
\begin{pmatrix}
a_{\rm pf} \\ a_{\rm pb}
\end{pmatrix}
= -i \sqrt{\Gamma_{\rm ep}}
\begin{pmatrix}
b_{\rm pf} \\ b_{\rm pb}
\end{pmatrix}. \label{ap_matrix}
\end{equation} 

Upon inverting the matrix in Eq.~\ref{ap_matrix}, we arrive at detuning-dependent solutions for the intracavity pump fields, which are given as

\begin{widetext}
\begin{eqnarray}
a_{\rm pf}(\Delta) &=& \frac{-i \sqrt{\Gamma_{\rm ep}}}{(i \Delta - \Gamma_{\rm tp}/2)^2 + \beta_p^2} \lbrace (i \Delta - \Gamma_{\rm tp}/2) b_{\rm pf} - i \beta_p e^{i \phi_{\beta_p}} b_{\rm pb} \rbrace, \label{apf_dualpumps} \\
a_{\rm pb}(\Delta) &=& \frac{-i \sqrt{\Gamma_{\rm ep}}}{(i \Delta - \Gamma_{\rm tp}/2)^2 + \beta_p^2} \lbrace - i \beta_p e^{-i \phi_{\beta_p}} b_{\rm pf} + (i \Delta - \Gamma_{\rm tp}/2) b_{\rm pb} \rbrace . \label{apb_dualpumps}
\end{eqnarray}
\end{widetext}

Equations \ref{apf_dualpumps} \& \ref{apb_dualpumps} describe the intracavity pump fields for the general case of two counter-propagating input pump waves (see Fig.~\ref{physical_picture}) entering the device. However, all results presented in the main text are obtained with a single forward-propagating input pump wave ($b_{pb} = 0$), such that the intracavity pump fields may be expressed in the following simpler forms, 

\begin{widetext}
\begin{eqnarray}
a_{\rm pf}(\Delta) &=& \frac{-i \sqrt{\Gamma_{\rm ep}}}{(i \Delta - \Gamma_{\rm tp}/2)^2 + \beta_p^2} \lbrace (i \Delta - \Gamma_{\rm tp}/2) b_{\rm pf} \rbrace, \label{apf_singlepump} \\
a_{\rm pb}(\Delta) &=& \frac{-i \sqrt{\Gamma_{\rm ep}}}{(i \Delta - \Gamma_{\rm tp}/2)^2 + \beta_p^2} \lbrace - i \beta_p e^{-i \phi_{\beta_p}} b_{\rm pf} \rbrace.  \label{apb_singlepump}
\end{eqnarray}
\end{widetext}

\section{Pair correlation theory plots for different waveguide-microdisk coupling rates}
Here, we describe the method used to generate the theoretically predicted pair correlation waveforms that appear in Fig. 5 of the main text. In Fig. 5(a)-(c) of the main text, we plot the experimental response of the pair correlations under three coupling regimes, $\Gamma < \beta$, $\Gamma \approx \beta$, and $\Gamma > \beta$, respectively, along with their corresponding theory plots in Fig. 5(d)-(f). Here, $\Gamma$ and $\beta$ represent the total photon decay rates and modal coupling rates, averaged between the signal and idler cavity modes.

\begin{figure*}[ht]
\begin{center}
\includegraphics[scale = 1]{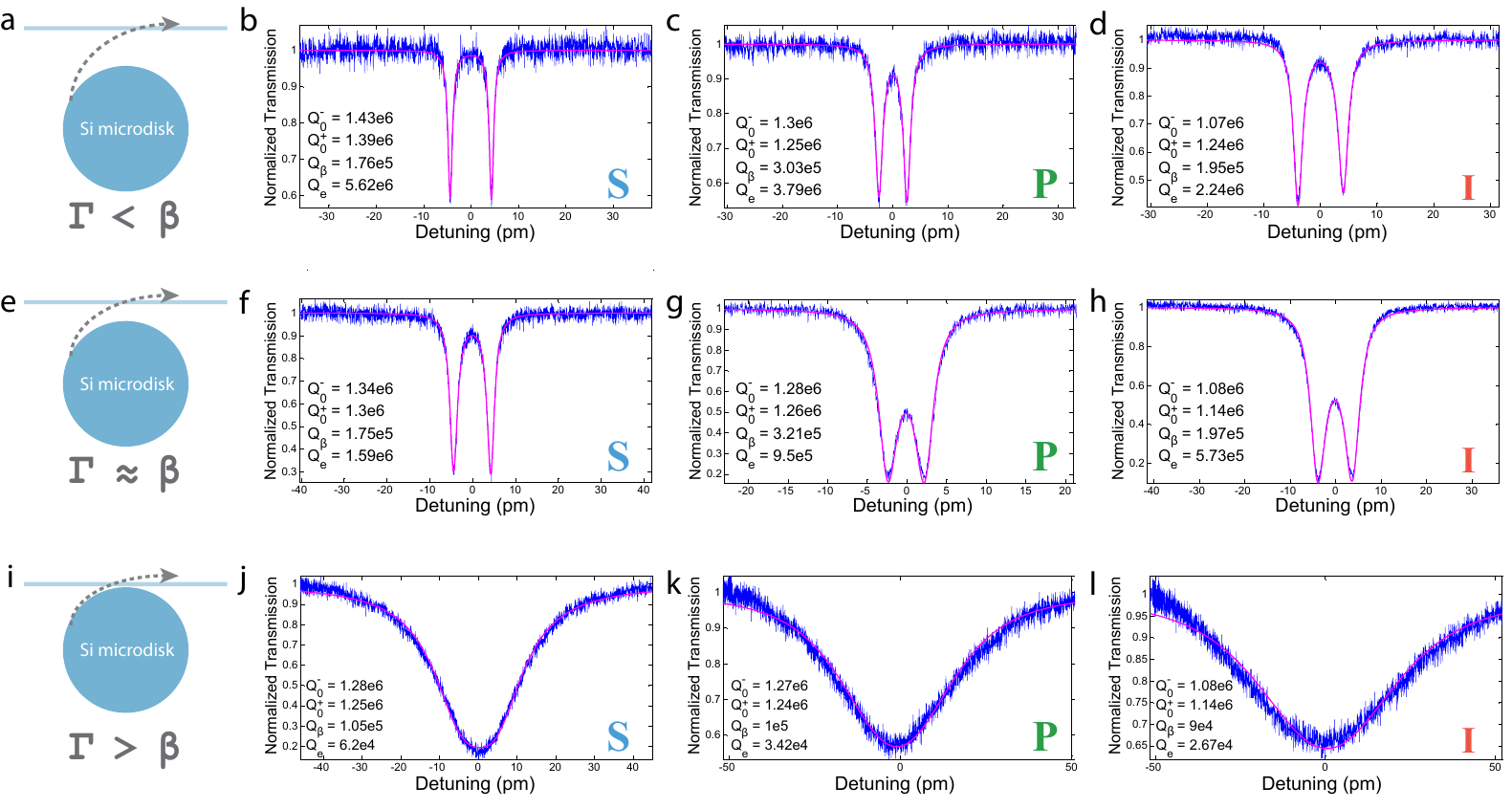}
\caption{\small Fits of the measured cavity transmission spectra for different coupling configurations. Optical fitting is used to determine the intrinsic photon decay rates ($\Gamma_0 = \omega_0/Q_0$), external coupling rates ($\Gamma_e = \omega_0/Q_e$), and modal coupling rates ($\beta = \omega_0/2Q_{\beta}$) for the three coupling regimes. (a) Illustration depicting the gap between the optical waveguide and microdisk in the $\Gamma < \beta$ case, along with corresponding fits of the signal (b), pump (c), and idler (d) transmission profiles. (e) Illustration depicting the waveguide-microdisk gap for the $\Gamma \approx \beta$ case, along with corresponding fits of the signal (f), pump (g), and idler (h) transmission profiles.(i) Illustration depicting the waveguide-microdisk gap for the $\Gamma > \beta$ case, along with corresponding fits of the signal (j), pump (k), and idler (l) transmission profiles. }
\label{MeasuredQBeta}
\end{center}
\end{figure*}
The three coupling regimes are experimentally accessed by varying the waveguide-microdisk gap, as is schematically shown in Fig.~\ref{MeasuredQBeta}. For each coupling case, we perform fits (shown in magenta) of the signal (s), pump (p), and idler (i) transmission profiles (shown in blue) in order to extract the intrinsic photon decay rates ($\Gamma_{\rm 0m} = \omega_{\rm 0m}/Q_{\rm 0m}$, where m = p,s,i), external coupling rates ($\Gamma_{\rm em} = \omega_{\rm 0m}/Q_{\rm em}$), and modal coupling rates ($\beta_{m} = \omega_{\rm 0m}/2 Q_{\beta_m}$). The measured cavity quantities are then inserted into Eq.~\ref{N}-\ref{c3}, which defines all of the constants relevant to the pair correlation theory (see Eq.~\ref{paircorr_ff}-\ref{paircorr_bb}). By incorporating the measured cavity quantities and laser-cavity detuning into Eq.~\ref{apf_singlepump} \& \ref{apb_singlepump}, we compute the relative phase and amplitude of the intracavity pump fields, such that the theory describing the pair correlations would be completely defined (no free parameters) with knowledge of the spatial phases associated with the orientation of the standing-wave mode patterns. As we do not possess the necessary experimental apparatus to measure the spatial phases that determine $\phi_{\beta}$, we consequently leave it as a free parameter in our theory.

\begin{figure*}[ht!]
\begin{center}
\includegraphics[scale = 0.80]{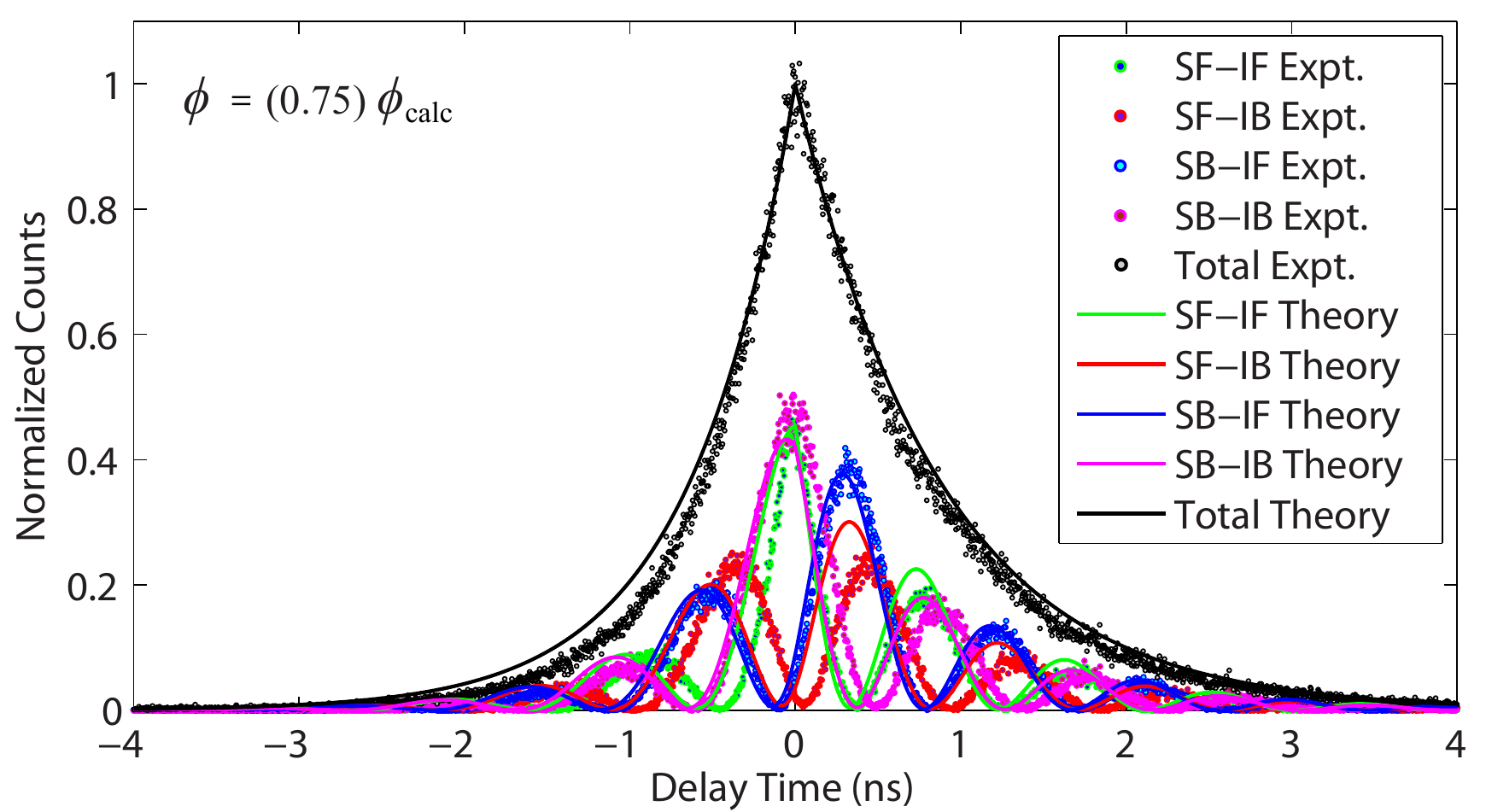}
\caption{\small Comparison between pair correlation theory and experimental results, with phase as the only free parameter. The cross-correlation waveforms (without background subtraction), measured between signal forward - idler forward (SF-IF) (green dots), signal forward - idler backward (SF-IB) (red dots), signal backward - idler forward (SB-IF) (blue dots), and signal backward - idler backward (SB-IB) (magenta dots) are plotted, along with their corresponding theoretically predicted pair correlation waveforms (solid lines). Inset denotes adjustment from theoretically calculated relative phase between intracavity pump modes.}
\label{undercoupled_theory_ideal}
\end{center}
\end{figure*}

In Fig.~\ref{undercoupled_theory_ideal}, we directly compare the pair correlation theory (shown as solid lines) to the experimental results (shown as dots) that appear in Fig. 5(a) of the main text. The theoretical curves are generated by incorporating the measured photon decay and coupling properties of the microdisk (see Fig.~\ref{MeasuredQBeta}(b)-(d)) into Eq.~\ref{paircorr_ff}-\ref{paircorr_bb}, and then varying the relative phase, $\phi$. We start with the relative phase computed from the intracavity pump fields (Eq.~\ref{apf_singlepump} \& \ref{apb_singlepump}), $\phi_{\rm calc}$, and then vary the phase until the best qualitative match among all the waveforms is reached. Here, it turns out that the best match results from using $\phi = 0.75\phi_{\rm calc}$. Consequently, these are the theoretical curves presented in Fig. 5(d) of the main text, which provide a qualitative comparison to the experimental data contained in Fig. 5(a) ($\Gamma < \beta$ case).

\begin{figure*}[ht!]
\begin{center}
\includegraphics[scale = 0.80]{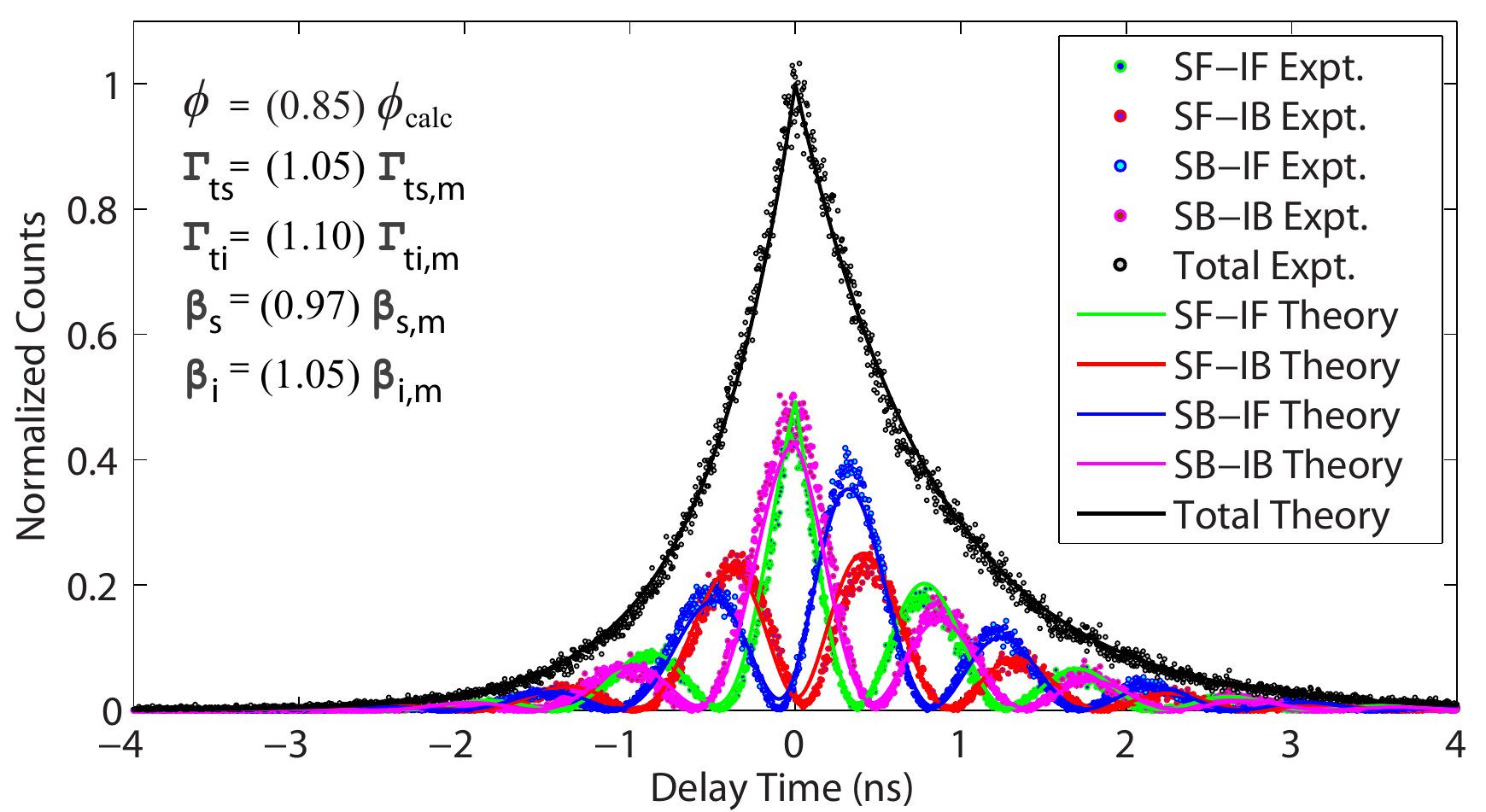}
\caption{\small Comparison between the pair correlation theory and experimental results, with the best fit achieved by small adjustments to multiple cavity-dependent quantities. The cross-correlation waveforms (without background subtraction), measured between signal forward - idler forward (SF-IF) (green dots), signal forward - idler backward (SF-IB) (red dots), signal backward - idler forward (SB-IF) (blue dots), and signal backward - idler backward (SB-IB) (magenta dots) are plotted, along with their corresponding theoretically predicted pair correlation waveforms (solid lines). Inset denotes adjustments from theoretically calculated relative phase between intracavity pump modes and measured (m) cavity properties.}
\label{undercoupled_theory_nonideal}
\end{center}
\end{figure*}
The theory curves in Fig.~\ref{undercoupled_theory_ideal} capture the essential behavior of the experimental results and have reasonable agreement. However, we note that some deviation is to be expected, given that our theory lacks certain details. For instance, the theory assumes that all the doublets are symmetric, which is not exactly the case for our device (see. Fig.~\ref{MeasuredQBeta}). The doublet asymmetry implies that there will be some discrepancy between the intracavity amplitudes and phases when comparing against the perfectly symmetric case. Additionally, we have not included effects due to two-photon absorption (TPA)/free-carrier absorption (FCA), which may have a small contribution to the deviation seen in Fig.~\ref{undercoupled_theory_ideal}. In particular, TPA/FCA may explain why the theoretically predicted photon decay envelope (black solid line) in Fig.~\ref{undercoupled_theory_ideal} is slightly broader than the corresponding experimental result (black dots). Given these non-idealities, we found it informative to test how good the match could be if certain measured parameters were allowed to be tuned by a small amount in the theory. In doing so, we have produced the theoretical curves shown in Fig.~\ref{undercoupled_theory_nonideal}, where the changes to measured values are listed in the inset. Here, the agreement between theory and experiment is extremely good, further strengthening the assertion that our theory describes the predominant physics of the system. However, we stress that these theoretical curves \textit{have not} been included in the main text, because they incorporate small changes to measured values. 

Thus far, we have described the manner in which the theory plots were obtained for the $\Gamma < \beta$ case. The same procedure was applied to the $\Gamma \approx \beta$ and $\Gamma > \beta$ cases, for which we used the measured values from Fig.~\ref{MeasuredQBeta}(f)-(h) and Fig.~\ref{MeasuredQBeta}(j)-(l), respectively. Additionally, in both of these cases the relative phase values that gave the best match between theory and experiment were $\phi_{\rm calc}$, the value computed directly from the intracavity pump fields (Eq.~\ref{apf_singlepump} \& \ref{apb_singlepump}), without any additional tuning. 

\section{Balancing the optical energy between intracavity pump modes}

\begin{figure*}[ht!]
\begin{center}
\includegraphics[scale = 1.65]{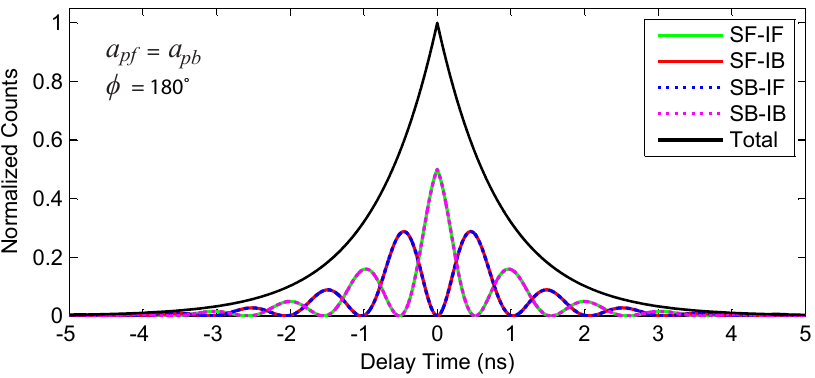}
\caption{\small Theoretical prediction of pair correlation waveforms with balanced intracavity pump energies and $\phi = 180^o$. For this special case, the correlation waveforms from the same path classification (co-propagating vs. counter-propagating) become indistinguishable. For simplicity, we have chosen to set the photon decay rates and modal coupling rates for the pumps, signal and idler cavity modes to be the same.}
\label{equal_pumps_phase180}
\end{center}
\end{figure*}

Here, we consider the implications of balancing the optical energy stored between the counter-propagating intracavity pump modes. In general, the intracavity pump modes will have unique values of stored optical energy ($f \neq b,$ where $f \equiv |a_{\rm pf}(\Delta)|^2$ and $b \equiv |a_{\rm pb}(\Delta)|^2$), as determined by the photon decay rates and modal coupling rates of the cavity, along with the laser-cavity detuning (see Eq.~\ref{apf_singlepump} \& \ref{apb_singlepump}). This is reflected in the pair correlation theory (see Eq.~\ref{paircorr_ff}-\ref{paircorr_bb}) by allowing $f$ and $b$ to assume any arbitrary values. However, in the special case that $f = b$, and after some minor algebra, the pair correlation functions become,

\begin{widetext}
\begin{eqnarray} 
p_{c,ff}(\tau) = 
	\begin{cases}
	  |f|^2 N e^{\Gamma_{\rm ti} \tau} |\lbrace c_0 e^{-i \phi} - c_1  \rbrace \cos{(\beta_i \tau)} + \lbrace c_2 e^{-i \phi} + c_3 \rbrace \sin{(\beta_i \tau)}|^2 & \quad (\tau < 0), \\
    |f|^2 N e^{-\Gamma_{\rm ts} \tau} |\lbrace c_0 e^{-i \phi} - c_1 \rbrace \cos{(\beta_s \tau)} - \lbrace c_3 e^{-i \phi} + c_2 \rbrace \sin{(\beta_s \tau)}|^2 & \quad (\tau \ge 0), \label{paircorr_ff_equal}
    \end{cases} \\ \nonumber \\
p_{c,fb}(\tau) = 
	\begin{cases}
	 |f|^2 N e^{\Gamma_{\rm ti} \tau} |\lbrace c_2 e^{-i \phi} + c_3 \rbrace \cos{(\beta_i \tau)} - \lbrace c_0 e^{-i \phi} - c_1 \rbrace \sin{(\beta_i \tau)}|^2 & \quad (\tau < 0), \\
     |f|^2 N e^{-\Gamma_{\rm ts} \tau} |\lbrace c_2 e^{-i \phi} + c_3 \rbrace \cos{(\beta_s \tau)} - \lbrace c_1 e^{-i \phi} - c_0 \rbrace \sin{(\beta_s \tau)}|^2 & \quad (\tau \ge 0), \label{paircorr_fb_equal}
    \end{cases} \\ \nonumber \\
p_{c,bf}(\tau) = 
	\begin{cases}
	 |f|^2 N e^{\Gamma_{\rm ti} \tau} |\lbrace c_2 e^{-i \phi} + c_3 \rbrace \cos{(\beta_i \tau)} - \lbrace c_0 e^{-i \phi} - c_1 \rbrace \sin{(\beta_i \tau)}|^2 & \quad (\tau < 0), \\
    |f|^2 N e^{-\Gamma_{\rm ts} \tau} |\lbrace c_2 e^{-i \phi} + c_3 \rbrace \cos{(\beta_s \tau)} - \lbrace c_1 e^{-i \phi} - c_0 \rbrace \sin{(\beta_s \tau)}|^2 & \quad (\tau \ge 0), \label{paircorr_bf_equal}
    \end{cases} \\ \nonumber \\
p_{c,bb}(\tau) = 
	\begin{cases}
	  |f|^2 N e^{\Gamma_{\rm ti} \tau} |\lbrace c_0 e^{-i \phi} - c_1 \rbrace \cos{(\beta_i \tau)} + \lbrace c_2 e^{-i \phi} + c_3 \rbrace \sin{(\beta_i \tau)}|^2 & \quad (\tau < 0), \\
    |f|^2 N e^{-\Gamma_{\rm ts} \tau} |\lbrace c_0 e^{i \phi} - c_1 b \rbrace \cos{(\beta_s \tau)} - \lbrace c_3 e^{-i \phi} + c_2 \rbrace \sin{(\beta_s \tau)}|^2 & \quad (\tau \ge 0), \label{paircorr_bb_equal}
    \end{cases} 
\end{eqnarray} 
\end{widetext}

\noindent where we have made use of the identity $|n_1 \pm n_2 e^{i \phi}|^2 = |n_1 \pm n_2 e^{-i \phi}|^2$ ($n_1$ and $n_2$ are constants). Thus, we see that when the optical energy stored between the intracavity pump modes is exactly balanced, the pair correlation waveforms from the same path classification (co-propagating vs. counter-propagating) become indistinguishable. A plot of the theory describing this phenomenon may be seen in Fig.~\ref{equal_pumps_phase180}, where we have set the photon decay rates and modal coupling rates between the pump, signal and idler cavity modes to be equivalent. 

\begin{figure*}[ht!]
\begin{center}
\includegraphics[scale = 1.65]{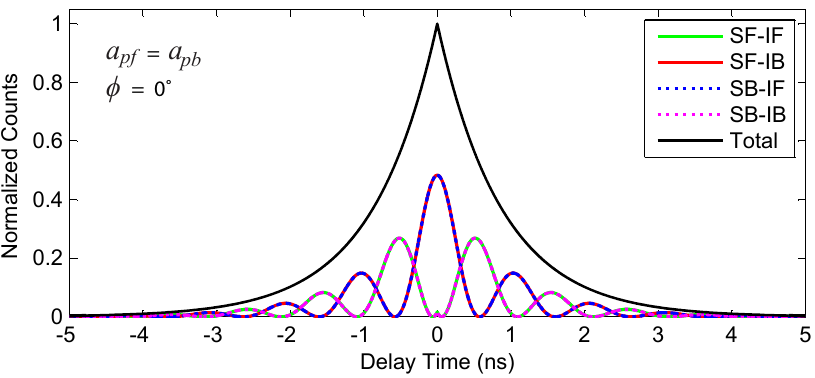}
\caption{\small Theoretical prediction of pair correlation waveforms with intracavity pump energies and $\phi = 0^o$. The pair correlations are flipped with respect to the $\phi = 180^o$ case. For simplicity, we have chosen to set the photon decay rates and modal coupling rates for the pumps, signal and idler cavity modes to be the same.}
\label{equal_pumps_phase0}
\end{center}
\end{figure*}

Here, we clearly see that the pair correlations for the forward-forward (SF-IF) and backward-backward (SB-IB) co-propagating path configurations are mutually indistinguishable, as well as the pair correlations associated with the forward-backward (SF-IB) and backward-forward (SB-IF) counter-propagating path configurations. As mentioned in the main text, this phenomenon is a consequence of the system possessing an internal mirror symmetry (with respect to the photon generation processes) upon exact balancing of the intracavity pump energies.

In Fig.~\ref{equal_pumps_phase180}, we see that the correlations belonging to the co-propagating state classification are maximally correlated at zero delay-time, whereas the opposite occurs for the correlations associated with the counter-propagating state classification. As discussed in the main text, by varying the relative pump phase, we can control the quantum interference within the device, such that this behavior is flipped between state classifications. With $\phi = 0^o$, as shown in Fig.~\ref{equal_pumps_phase0}, the correlations waveforms from the same state classification are now perfectly transposed in comparison to Fig.~\ref{equal_pumps_phase180}. We note that the small hump appearing in the correlation waveforms corresponding to the co-propagating states (see Fig.~\ref{equal_pumps_phase0}) is a consequence of the coupling vs total photon decay rate. This feature can be made arbitrarily small by increasing the rate of coupling compared to the rate of decay (as we have done in theoretical investigations not presented here). In fact, the presence or absence of this feature is a manifestation of the distinguishability between alternative two-photon pathways within the system. When the coupling rate is comparable to the photon decay rate, we gain some knowledge about whether a photon originally came from the forward or backward traveling mode. However, when the coupling rate is much greater than the photon decay rate, then the 'which-path' information is totally removed from the system, allowing for perfect interference visibility and complete removal of the hump.
 
\section{Response of pair correlations to laser-cavity detuning around resonance}

\begin{figure*}[ht!]
\begin{center}
\includegraphics[scale = 1]{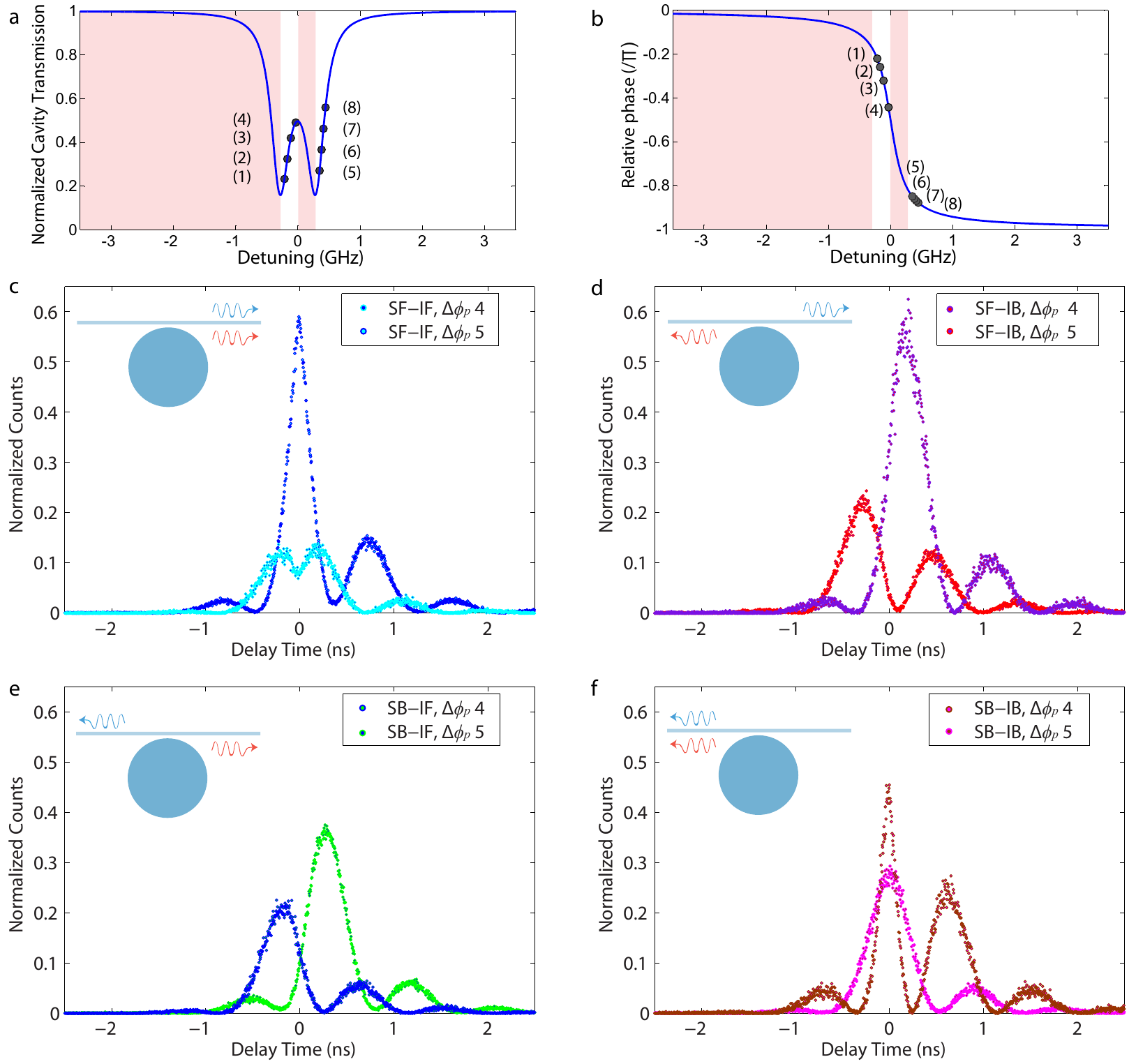}
\caption{\small Response of the pair correlations to laser-cavity detuning across the pump resonance. Theoretical plots of (a) the pump transmission profile and (b) relative phase between counter-propagating pump modes with the labels, (1)-(8) denoting the detuning values $\Delta$ = (-0.21, -0.16, -0.11, -0.06, 0.35, 0.38, 0.41, 0.44) GHz, respectively. The light-red shading indicates the unstable locking regions. Cross-correlation waveforms (without background subtraction), measured between (c) signal forward - idler forward (SF-IF), (d) signal forward - idler backward (SF-IB), (e) signal backward - idler forward (SB-IF), and (f) signal backward - idler backward (SB-IB) for detuning values (4) and (5) only. As seen in (b), the transition from (4) to (5) tunes the system through resonance and consequently induces the largest phase shift. Insets depict path configuration.}
\label{DetAcrossRes}
\end{center}
\end{figure*}
Here, we provide detailed versions of a subset of the pair correlations appearing in Fig. 6 of the main text. The experimental results in that figure were obtained by varying the laser-cavity detuning about resonance (as indicated by the laser locking points in Fig.~\ref{DetAcrossRes}(a), which was copied from Fig. 6(a) in the main text), which induces a large relative phase shift between the counter-propagating intracavity pump modes, as shown in Fig.~\ref{DetAcrossRes}(b) (which was copied from Fig. 6(b) of the main text). Although all eight points are important in characterizing the pump-induced interference phenomena, the most pronounced changes occur between the detuning values labeled as (4) and (5) (see Fig.~\ref{DetAcrossRes}(a),(b)), because this transition includes tuning through the resonance frequency of the pump cavity mode. Here, we provide detailed versions of the response of the pair correlations for each path configuration as the carrier frequency of the pump laser is swept from negative to positive detuning.  

\begin{figure*}[ht!]
\begin{center}
\includegraphics[scale = 1.05]{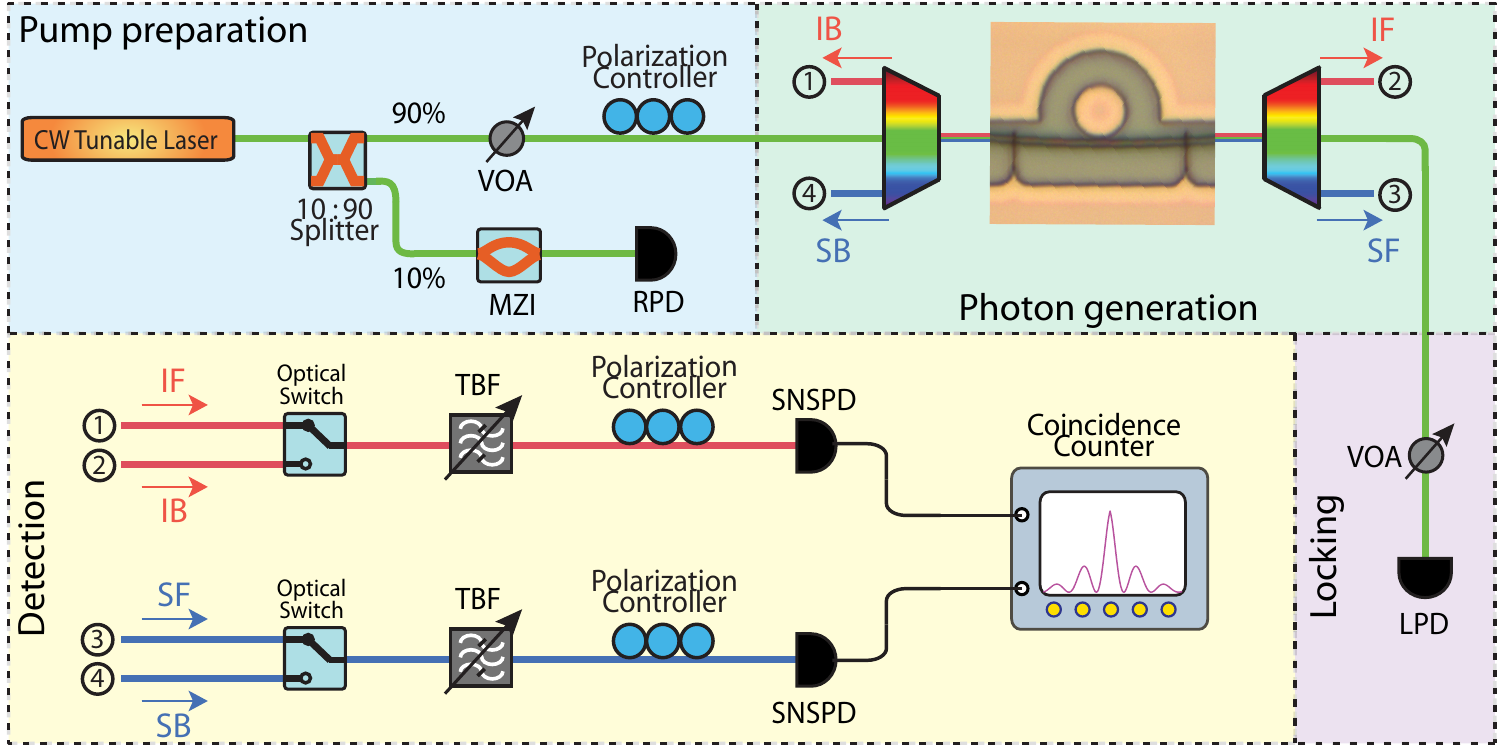}
\caption{\small Schematic of the experimental setup used to obtain all results presented in the main text. A forward-propagating pump wave is launched into optical fiber and is split using a 10/90 directional coupler. A small portion (10\%) of the pump light is passed through a Mach-Zehnder interferometer (MZI) and detected with a fast reference photodiode (RPD) in order to accurately calibrate the fitting of optical resonances. The primary portion (90\%) of the pump light passes through a variable optical attenuator (VOA) to control the input optical power, a fiber polarization controller (FPC) to align the polarization state to that of the quasi-transverse magnetic mode family within the microdisk, and a course-wavelength division multiplexing (CWDM) multiplexer (MUX) before evanescently coupling into the microdisk. The transmitted pump light is separated from the single photon channels by a demultiplexer (DEMUX) and detected with a fast locking photodiode (LPD), in order to lock the laser to the cavity. Within the microdisk, signal (s) and idler (i) photon pairs are created between the coherently coupled forward (f) and backward (b) traveling-wave modes via spontaneous four-wave mixing, such that four single photon pathways are established: signal-forward (SF), signal-backward (SB), idler-forward (IF), and idler-backward (IB). The four single photon pathways are separated by demultiplexing and sent to optical switches. Photons exiting the switches are then passed through narrowband tunable bandpass filters (TBF) which remove the Raman noise generated in the delivery optical fiber. The photon pairs are ultimately detected using superconducting nanowire single photon detectors (SNSPD) with state-of-the-art timing resolution. The photon arrivals are then processed with a coincidence counter in the time-tagged configuration, such that the absolute arrival time of every photon detection event is recorded. }  
\label{ExptSetup}
\end{center}
\end{figure*}


\newpage
\section{Experimental setup}
Here, we provide a schematic of the experimental setup (see Fig.~\ref{ExptSetup}) used to obtain all of the results appearing in the main text. A discussion of the experimental techniques and procedures may be found in Appendix B of the main text; an abridged version of which may be found in the caption of Fig.~\ref{ExptSetup}.


\end{document}